\documentclass[twocolumn,aps,superscriptaddress]{revtex4-1}
\usepackage{palatino}
\usepackage{natbib}
\usepackage[latin1]{inputenc}
\usepackage{epsf}
\usepackage{amsmath,amssymb}
\usepackage{latexsym}
\usepackage{calc}
\usepackage{color}
\usepackage{shadow}
\usepackage{epsfig}

\newcommand{\ben}{\begin{equation}}
\newcommand{\een}{\end{equation}}
\newcommand{\bea}{\begin{eqnarray}}
\newcommand{\eea}{\end{eqnarray}}

\newcommand{\nn}{\nonumber}

\def\ext{_{\rm ext}}

\def\br{{\bf r}}
\def\bR{{\bf R}}

\def\dulR{{\underline{\underline{\bf R}}}}
\def\dulr{{\underline{\underline{\bf r}}}}

\def\dulS{{\underline{\underline{\sigma}}}}
\def\duls{{\underline{\underline{\bf s}}}}

\def\na{{\nabla}}

\begin{document}
  \title{Correlated electron-nuclear dynamics: Exact factorization of the molecular wavefunction} 
  \author{Ali Abedi}
\affiliation{Max-Planck Institut f\"ur Mikrostrukturphysik, Weinberg 2, D-06120 Halle, Germany}
\affiliation{European Theoretical Spectroscopy Facility (ETSF)}
\author{Neepa T. Maitra} 
\affiliation{Department of Physics and Astronomy, Hunter College and the City University of New York, 695 Park Avenue, New York, New York 10065, USA}
\author{E.K.U. Gross}
\affiliation{Max-Planck Institut f\"ur Mikrostrukturphysik, Weinberg 2, D-06120 Halle, Germany}
\affiliation{European Theoretical Spectroscopy Facility (ETSF)}

  \date{\today}
  \begin{abstract}
It was recently shown~\cite{AMG10} that the complete wavefunction for
a system of electrons and nuclei evolving in a time-dependent external
potential can be exactly factorized into an electronic wavefunction and
a nuclear wavefunction. The concepts of an exact time-dependent
potential energy surface (TDPES) and exact time-dependent vector potential emerge
naturally from the formalism. Here we present a detailed description of the
formalism, including a full derivation of the equations that the
electronic and nuclear wavefunctions satisfy. We demonstrate the
relationship of this exact factorization to the traditional
Born-Oppenheimer expansion. A one-dimensional model of the H$_2^+$
molecule in a laser field shows the usefulness of the exact TDPES in
interpreting coupled electron-nuclear dynamics: we show how features
of its structure indicate the mechanism of dissociation. We compare
the exact TDPES with potential energy surfaces from the time-dependent
Hartree-approach, and also compare traditional Ehrenfest dynamics with
Ehrenfest dynamics on the exact TDPES.
 \end{abstract}
 \maketitle 
 \section{Introduction}
The interplay of nuclear and electronic dynamics in the presence of
time-dependent external fields leads to fascinating phenomena,
especially beyond the perturbative regime, e.g.  photo-induced
molecular dissociation, charge-resonance enhanced ionization, control
of electron localization, electron-hole migration after photo-excitation, to name a
few~\cite{BK03,Marangos04,Kling04, DP07, Rozzi11}.  
 The
exact solution of the time-dependent Schr\"odinger equation (TDSE) is
currently out of computational reach except for the very simplest of
molecules~\cite{CCZB96}, such as H$_2^+$, so usually approximate
methods are used. Typically, (but not always, see
Refs.~\cite{HBNS06,Martinez,Martin07,paramonov05}), these methods treat the nuclei
classically as point charges with electron-nuclear coupling given by
Ehrenfest dynamics, or surface-hopping~\cite{TIW96p}; a topical
application is to model photochemical processes~\cite{TTRF08,PDP09},
for example, in solar cells, to study the (field-free) dynamics ensuing after
an initial electronic excitation. 
Indeed several examples have shown that the predicted
electron-hole migration can depend critically on the description of
the nuclear motion and how it is correlated with the electronic
dynamics~(see Ref. \cite{DP07,Rozzi11} and references within).
Apart from enabling calculations on more than the simplest systems possible, these methods 
provide much intuition, in particular
through the central concept of the potential energy surface
(PES). Indeed, the very idea itself of surface-hopping would
not exist without the notion of a landscape of coupled PESs. Dressed molecular potentials such as light-induced molecular
potentials (LIMPS)~\cite{BS81} have proved valuable in
understanding processes such as bond-softening, stabilization against
dissociation, etc. where the laser field induces avoided crossings
between PESs.  
Approximate time-dependent potential energy surfaces (TDPES) were
introduced by Kono~\cite{KSTK04} as instantaneous eigenvalues of the
electronic Hamiltonian, and have proven extremely useful in the
interpretation of system-field phenomena, as have the quasi-static or
phase-adiabatic PES's used recently to interpret electron localization in dissociative ionization~\cite{KSIV11}. Recent work of
Cederbaum~\cite{C08} introduced a TDPES in a different way, by
generalizing the Born-Oppenheimer approximation to include
time-dependent external potentials.  In short, the PES is perhaps the
most central concept in our understanding of molecular motion.

In a recent Letter~\cite{AMG10}, we showed that an {\it exact} TDPES
may be defined, via a rigorous separation of electronic and nuclear
motion by introducing an {\it exact} factorization of the full
electron-nuclear wavefunction. The idea of an exact factorization was first introduced by Hunter~\cite{Hunter} for the static case. 
He also deduced the exact equation of motion for the nuclear factor. The equation of motion for the electronic wavefunction was first
given by Gidopoulos and Gross~\cite{GG05} for the time-independent case. Both in the static and in the time-dependent case the
factorization leads to an exact definition of the PES, and also of the
Berry vector potential. What is particularly interesting about the
vector potential is that Berry-Pancharatnam phases~\cite{Berry} are
usually interpreted as arising from some {\it approximation} where a
system is decoupled from ``the rest of the world'', thereby making the
system Hamiltonian dependent on some ``environmental'' parameters. For
example, in the static BO approximation, the electronic Hamiltonian
depends parametrically on nuclear positions, and when the molecular
wavefunction is approximated by a single product of a nuclear
wavefunction and an eigenstate of the electronic Hamiltonian, the
equation of motion for the former contains a Berry vector potential. The question whether the BO Berry phase survives in the exact
treatment was first discussed in Ref.~\cite{GG05} for the static case and in Ref.~\cite{AMG10} for the time-dependent case. 

In the present paper, we provide the detailed derivation of the
formalism of Ref.~\cite{AMG10} (Section~\ref{sec:exactfact}), analyse
features of the exact electron-nuclear coupling terms in general
(Section~\ref{sec:threeterms}), including their relationship to
couplings in the traditional Born-Oppenheimer expansion, and then
study the TDPES for the specific case of a model H$_2^+$ molecule in
an oscillating electric field (Section~\ref{sec:H2plus}). The
remainder of this introduction serves to set up the problem at hand,
and to remind the reader of the Born-Oppenheimer treatment of the
electron-nuclear system.

 \subsection{The Hamiltonian}
In this section we establish notation and define the Hamiltonian for the combined system of electrons and nuclei. The coordinates of the $N_e$ electrons are 
collectively denoted by $\dulr~\duls$ where $\dulr \equiv \{r_j\}$ and $\duls \equiv \{s_j\}, j=1...N_e$, represent electronic spatial and spin coordinates, 
respectively. The $N_n$ nuclei have masses $M_1...M_{N_n}$ and charges $Z_1...Z_{N_n}$ and coordinates collectively denoted by $\dulR~\dulS$ where 
$\dulR \equiv \{R_{\alpha}\}$ and $\dulS \equiv \{\sigma_{\alpha}\}, \alpha = 1...N_n$, represent nuclear spatial and spin coordinates, respectively. Furthermore,
we consider the system is under the influence of some time-dependent external scalar field. The system is described, non-relativistically, by the Hamiltonian 
 \ben
   \hat{H} = \hat{H}_{BO}(\dulr,\dulR) + \hat{V}\ext^e(\dulr,t)+\hat{T}_n(\dulR) + \hat{V}\ext^n(\dulR,t),
\label{eq:H}
 \een
 where $\hat H_{BO}(\dulr,\dulR) $ is the familiar Born-Oppenheimer electronic Hamiltonian,
 \ben
 \hat{H}_{BO} = \hat{T}_e(\dulr) + \hat{W}_{ee}(\dulr) + \hat{W}_{en}(\dulr,\dulR) + \hat{W}_{nn}(\dulR).  
 \label{eq:tradBO}
 \een

 The subscripts ``e'' and ``n'' refer to electrons and nuclei, respectively, and atomic units are used throughout ($e^2 = \hbar = m_e = 1$). Here
 \ben
 \label{Te}
 \hat{T}_e=-\sum_{j=1}^{N_e} \frac{1}{2}\na_j^2
 \een
 and
 \ben
 \label{Tn}
 \hat{T}_n=-\sum_{\alpha=1}^{N_n}\frac{1}{2M_{\alpha}}\na_{\alpha}^2
 \een
 denote the kinetic-energy operators of the electrons and nuclei, respectively. All external scalar potentials 
on the system (e.g. electric fields) are represented by
\ben
 \label{ext-n}
 \hat{V}\ext^n = \sum_{\alpha}^{N_n} v\ext^n(\bR_{\alpha},t),
 \een
and
  \ben
 \label{ext-e}
 \hat{V}\ext^e = \sum_{j}^{N_e} v\ext^e(\br_j,t),
 \een
The particle-particle Coulomb interactions have the form:
 \ben
 \label{nnc}
 \hat{W}_{nn} = \frac{1}{2} \sum_{\substack{\alpha,\beta=1\\\alpha\neq\beta}}^{N_n} \frac{Z_{\alpha}Z_{\beta}}{\vert \bR_{\alpha} - \bR_{\beta} \vert},
 \een

 \ben
 \label{eec}
 \hat{W}_{ee} = \frac{1}{2} \sum_{\substack{i,j=1\\i\neq j}}^{N_e} \frac{1}{\vert \br_i - \br_j \vert},
 \een
 
 \ben
 \label{enc}
 \hat{W}_{en} = - \sum_{j}^{N_e} \sum_{\alpha}^{N_n} \frac{Z_{\alpha}}{\vert \br_{j} - \bR_{\alpha} \vert}.
 \een

 The quantum mechanical equation of motion of such a system is given by the TDSE:
 \ben
 \label{eq:tdse}
 \hat{H}\Psi(\dulr~\duls,\dulR~\dulS,t) = i\partial_t \Psi(\dulr~\duls,\dulR~\dulS,t)
 \een

 The full electron-nuclear wavefunction,
 $\Psi(\dulr~\duls,\dulR~\dulS,t)$, that satisfies the
 TDSE~(\ref{eq:tdse}), contains the complete information on the
 system. As discussed in the introduction, it can be solved numerically
 only for very small systems of one or two electrons and nuclei and
 , moreover, $\Psi$ does not give access to PESs, which provide an intuitive
 understanding and interpretation of the coupled electron-nuclear dynamics.

 \subsection{The Born-Oppenheimer Approximation}

The Born-Oppenheimer (BO) approximation is among the most basic
 approximations in the quantum theory of molecules and solids. Consider the case when there is no external time-dependence in the Hamiltonian. The BO approximation relies on the fact that electrons typically move much faster than the
 nuclei; on the timescale of nuclear motion, the electrons
 ``instantly'' adjust to remain on the instantaneous eigenstate. This ``adiabatic approximation''
 allows us to visualize a molecule or solid as a set of nuclei moving
 on the PES generated by the electrons in a
 specific electronic eigenstate. The electronic Hamiltonian
 $\hat{H}_{BO}(\dulr,\dulR)$ depends parametrically on the nuclear
 positions, via the electron-nuclear Coulomb interaction.  That is,
 the stationary electronic Schr\"odinger equation is solved for each
 fixed nuclear configuration $\dulR~\dulS$,
\ben
 \label{eq:boe}
 \hat{H}_{BO}(\dulr,~\dulR)\phi^j_{\dulR~\dulS}(\dulr~\duls) = V^j_{BO}(\dulR~\dulS)\phi^j_{\dulR~\dulS}(\dulr~\duls) 
 \een
 yielding ($\dulR~\dulS$)-dependent eigenvalues $V^j_{BO}(\dulR~\dulS)$ and eigenfunctions $\phi^j_{\dulR~\dulS}$. The total molecular wavefunction, $\Psi_{BO}(\dulr~\duls,\dulR~\dulS)$,
is then approximated 
 as a product of the relevant electronic state, $\phi^j_{\dulR~\dulS}(\dulr~\duls)$, and a nuclear wavefunction $\chi^{BO}_{j\nu}(\dulR~\dulS)$ satisfying the corresponding
 BO nuclear Schr\"odinger equation 
 \bea
\nonumber
& \left(\sum_{\alpha=1}^{N_n}\frac{1}{2M_\alpha}(-i\nabla_\alpha+{\cal F}_{jj,\alpha}^{BO}(\dulR~\dulS))^2+\epsilon^j_{BO}(\dulR~\dulS)\right)\chi^{BO}_{j\nu}(\dulR~\dulS)&\\
 &= E \chi^{BO}_{j\nu}(\dulR~\dulS)&
 \label{eq:bon} 
\eea
 where
 \bea
 \label{eq:bopes}
& \epsilon^j_{BO}(\dulR~\dulS) = &\\
&\sum_{\duls}\left\langle\phi^j_{\dulR~\dulS} \right\vert\hat{H}_{BO}(\dulr, \dulR)+\sum_{\alpha}\frac{(-i\nabla_\alpha-{\cal F}_{jj,\alpha}^{BO})^2}{2M_{\alpha}}\left\vert \phi^j_{\dulR~\dulS}\right\rangle_\dulr &
\nonumber
 \eea
and
\ben
 \label{eq:bovp}
 {\cal F}_{jj,\alpha}^{BO}(\dulR~\dulS)=-i\sum_{\duls}\langle\phi^j_{\dulR~\dulS}\vert\nabla_\alpha\phi^j_{\dulR~\dulS}\rangle_\dulr\;.
 \een
where $\langle ..|..|..\rangle_\dulr$ denotes an inner product over all spatial electronic variables only.
The index $\nu$ of the nuclear wave function labels the vibrational/rotational eigenstate on the $j$th PES. The second term on the right 
of Eq.~\ref{eq:bopes} is often referred
to as the ``BO diagonal correction'' or ``adiabatic correction''. The
 potential energy surface $\epsilon^j_{BO}(\dulR~\dulS)$
is enormously important in molecular physics and quantum chemistry.
It is a central tool in the analysis and interpretation of molecular
absorption and emission spectra, experiments involving nuclear
motion, mechanisms of dissociation, energy-transfer, for
example. The nuclear dynamics on a {\it single} PES (sometimes called ``BO dynamics'') is obtained 
by using the Hamiltonian on the left of Eq.~(\ref{eq:bon}) in a time-dependent Schr\"odinger equation for a
time-dependent nuclear wavefunction $\chi(\dulR~\dulS,t)$. This corresponds to approximating the total 
molecular wavefunction by a time-dependent nuclear wavepacket multiplied with a static electronic BO state:
\ben
\Psi(\dulr~\duls,\dulR~\dulS,t) \approx \chi^{BO}(\dulR~\dulS,t)\phi^j_{\dulR~\dulS}(\dulr~\duls).
\een
The vector potential ${\cal F}_{jj,\alpha}^{BO}(\dulR~\dulS)$,
especially the Berry phase associated with it, $\oint{\cal
  F}_{jj,\alpha}^{BO}(\dulR~\dulS)\cdot d\dulR$, captures the essential
features of the behavior of a system with conical
intersections. Inclusion of the Berry phase can significantly shift
and re-order the energy eigenvalues of molecular roto-vibrational
spectra, as well as scattering cross-sections (although sometimes
undetected in experiments that measure integrated quantities, due to
cancellations between paths, see e.g. Refs.~\cite{M92,K03,R00,BALB10,A06} and references within).

It appears from the above discussion that in the traditional treatment
of molecules and solids the concepts of the PES
and the Berry phase arise as a consequence of the BO approximation.
Some of the most fascinating
phenomena of condensed-matter physics, like superconductivity, however, appear 
in the regime where the BO approximation is not valid; likewise typical photodynamical processes in molecules require going beyond the single-electronic-surface picture. This raises the
question: If one were to solve the Schr\"odinger equation of the full
electron-nuclear Hamiltonian exactly (i.e. beyond the BO
approximation) do the Berry phase and the potential energy surface
survive, with a possibly modified form,  and if so, how and where do they show up? What is their
relation to the traditional potential energy surface and Berry phase
in the BO approximation? Moreover, many interesting phenomena occur when molecules
or solids are exposed to time-dependent external field
e.g. lasers. Can one give a precise meaning to a {\it time-dependent}
potential energy surface and a time-dependent vector potential?

 Before answering the points raised above,
 focussing on the time-dependent case, we briefly discuss the Born-Oppenheimer {\it expansion} which solves the full TDSE~Eq.~(\ref{eq:tdse}) exactly for the coupled
 electron-nuclear system.
 
 \subsection{The Born-Oppenheimer Expansion}  
\label{sec:BOexp}
  The set of electronic eigenfunctions~$\{\phi^j_{\dulR~\dulS}(\dulr~\duls)\}$ calculated from Eq.~(\ref{eq:boe}) form a complete orthonormal set in the 
 electronic space for each fixed $\dulR~\dulS$ 
 \ben
 \label{eq:BO-EN}
 \sum_{\duls}\int d\dulr\phi^{l*}_{\dulR~\dulS}(\dulr~\duls)\phi^{j}_{\dulR~\dulS}(\dulr~\duls) = \delta_{lj}, 
 \een
 therefore the total time-dependent wavefunction of the system $\Psi(\dulr~\duls,\dulR~\dulS,t)$ can be expanded in that basis:
 \ben
 \label{eq:BO-Exp}
 \Psi(\dulr~\duls,\dulR~\dulS,t) = \sum_{j=1}^{\infty} \chi^{BO}_j(\dulR~\dulS,t)\phi^j_{\dulR~\dulS}(\dulr~\duls)\;.
 \een
 Here
 \ben
 \label{eq:BO-ExpCof}
 \chi^{BO}_j(\dulR~\dulS,t) = \sum_{\duls}\int d\dulr\phi^{j*}_{\dulR~\dulS}(\dulr~\duls)\Psi(\dulr~\duls,\dulR~\dulS,t)
 \een
 are the expansion coefficients which are functions of the nuclear degrees of freedom and time. 
Eq.~(\ref{eq:BO-Exp}) is the so-called BO expansion which
 is an {\it exact} representation of the complete molecular wavefunction due to the completeness of~$\{\phi^j_{\dulR~\dulS}(\dulr~\duls)\}$. It applies also to fully-time-dependent problems where $\Psi$ evolves under  external time-dependent potentials $\hat{V}\ext^e$. In practice, for numerically feasible calculations,
 approximations are introduced to limit the expansion to a small subset of~$\{\phi^j_{\dulR~\dulS}(\dulr~\duls)\}$.
 By inserting the expansion~(\ref{eq:BO-Exp}) into Eq.~(\ref{eq:tdse}), multiplying by $\phi^{j*}_{\dulR~\dulS}(\dulr~\duls)$ from the left, and integrating over the 
 electronic degrees of freedom, equations for the expansion coefficients $\chi^{BO}_j(\dulR~\dulS,t)$ are determined. One obtains:
\begin{widetext}
 \ben 
 \Big[\sum_{\alpha}\frac{1}{2M_{\alpha}}(-i\nabla_\alpha+{\cal F}^{BO}_{kk,\alpha})^2+ \hat{V}\ext^n+\epsilon_{BO}^k\Big]\chi^{BO}_k+\sum_{j\neq k} \Big[<\phi^k\vert\hat{V}\ext^e(t)\vert\phi^j> -\sum_{\alpha}\Lambda^{BO}_{kj,\alpha}\Big]\chi^{BO}_j= i\frac{\partial \chi^{BO}_k}{\partial t}\;.
 \een
Here  
 \ben
 \epsilon_{BO}^k(\dulR~\dulS,t) = \sum_{\dulS}\left\langle\phi^k_{\dulR~\dulS}\right\vert\hat{H}_{BO} + \hat{V}\ext^e+\sum_{\alpha}\frac{(-i\nabla_\alpha-{\cal F}^{BO}_{kk,\alpha})^2}{2M_{\alpha}}\left\vert \phi^k_{\dulR~\dulS}\right\rangle_{\dulr}
 \een
\end{widetext}
is the time-dependent scalar potential and is the $k$th generalized BO potential energy, generalized to account for the time-dependent external field (c.f. Eq.~(\ref{eq:bopes})). The terms
\ben
 \label{eq:NAC-BO}
 \Lambda^{BO}_{kj,\alpha}(\dulR) = \frac{1}{2M_{\alpha}}\Big[{\cal G}^{BO}_{kj,\alpha}(\dulR)+2{\cal F}^{BO}_{kj,\alpha}(\dulR)\cdot(i\nabla_{\alpha})\Big]
 \een
 are called the ``nonadiabatic couplings'', defined by~\cite{BH54, CBBO, Baer}:
 \bea
 \label{eq:NA-BO}
 {\cal F}^{BO}_{kj,\alpha}(\dulR) = -i<\phi^k_{\dulR~\dulS}\vert\nabla_{\alpha}\phi^j_{\dulR~\dulS}>\nonumber\\
 {\cal G}^{BO}_{kj,\alpha}(\dulR) = <\phi^k_{\dulR~\dulS}\vert\nabla^2_{\alpha}\phi^j_{\dulR~\dulS}>
 \eea

\section{Exact factorization of the time-dependent electron-nuclear wavefunction}
\label{sec:exactfact}
The BO expansion Eq.~(\ref{eq:BO-Exp}) yields the complete molecular
wavefunction exactly.  Instead of having an infinite sum of terms involving an infinite set of generalized PES's and non-adiabatic couplings, the question arises whether
it is possible to represent the complete, time-dependent, electron-nuclear wavefunction
exactly as a {\it single} product of an electronic wavefunction and a
nuclear wavefunction. In this section, we show that the
answer is yes.  We derive formally exact equations of motion for each
subsystem, out of which emerge rigorous definitions of a
time-dependent potential energy surface (TDPES) and a time-dependent
vector potential. 

Visually, the decomposition is similar in form to the single-surface
BO approximation, yet it is exact.  There is no assumption on the time
scale of the motions of each subsystem, i.e. unlike in the BO
approximation, we do not solve for the ``fast'' variables first
and then feed it into the equation for the
``slower'' variables. 
Instead,
the equations of motion for each subsystem are derived together, in a variational approach.
 The exact decomposition, contrary
to the BO separation, accounts for the full correlation
between the two subsystems, regardless of the mass and energy of the nuclear subsystem. In the following we
formalize the idea as a  theorem which we then prove. We discuss in detail the implications of this exact
decomposition.

\subsection{The exact factorization}
{\bf Theorem I. (a)} {\it The \underline{exact} solution of Eq.~(\ref{eq:tdse}) can be written as a single product} 
\begin{equation}
  \Psi(\dulr~\duls,\dulR~\dulS,t)=\Phi_{\dulR~\dulS}(\dulr~\duls,t)\chi(\dulR~\dulS,t)
\label{eq:product-ansatz}
\end{equation}
{\it where $\Phi_{\dulR~\dulS}(\dulr~\duls,t)$ satisfies the Partial Normalization Condition (PNC),} 
\ben
\sum_{\duls}\int d\dulr\vert\Phi_{\dulR~\dulS}(\dulr~\duls,t)\vert^2=1 \;,
\label{eq:pnc}
\end{equation}  
{\it for any fixed nuclear configuration, $\dulR~\dulS$, at any time $t$.} 

The PNC is critical in making this theorem
meaningful: Eq.~(\ref{eq:product-ansatz}) on its own would be rather
meaningless, because, for example, one could then simply just take
$\chi(\dulR~\dulS,t)\equiv 1$. In fact, one can come up with many different decompositions that satisfy Eq.~(\ref{eq:product-ansatz}) but that violate  the PNC Eq.~(\ref{eq:pnc}); it is the latter that makes the decomposition unique up to a gauge-like transformation, as we shall see shortly in Section~\ref{sec:thm2}. We will also see there that it is the PNC that  allows the interpretation of $\Phi_{\dulR~\dulS}(\dulr~\duls,t)$ as a conditional probability amplitude, and $\chi(\dulR~\dulS,t)$ as a marginal probability amplitude, leading to their identification as electronic and nuclear wavefunctions respectively.
First, we prove Part(a) of Theorem I. 

{\underline{\it Proof:}} Given $\Psi(\dulr~\duls,\dulR~\dulS,t)$, 
 the exact solution of the full TDSE~(\ref{eq:tdse}), we choose $\chi(\dulR~\dulS,t)$ and $\Phi_{\dulR~\dulS}(\dulr~\duls,t)$, at any instant in time, as
\bea
\label{eq:proof-Ia-phi}
&\chi(\dulR~\dulS,t) =  e^{iS(\dulR~\dulS,t)}\sqrt{\sum_{\duls}\int d\dulr\vert\Psi(\dulr~\duls,\dulR~\dulS,t)\vert^2}\;\;\\  &{\rm and} \nonumber\\
\label{eq:proof-Ia-chi}
&\Phi_{\dulR~\dulS}(\dulr~\duls,t) = \Psi(\dulr~\duls,\dulR~\dulS,t) / \chi(\dulR~\dulS,t)
\eea
where $S(\dulR~\dulS,t)$ is real. The PNC Eq.~(\ref{eq:pnc}) then follows immediately:
 \bea
\label{proof-Ia-pn}
\nonumber
 \sum_{\duls}\int d\dulr\vert\Phi_{\dulR~\dulS}(\dulr~\duls,t)\vert^2 &=& \frac{\sum_{\duls}\int d\dulr\vert\Psi(\dulr~\duls,\dulR~\dulS,t) \vert^2}{\vert\chi(\dulR~\dulS,t)\vert^2} \\
&=& \frac{\vert\chi(\dulR~\dulS,t)\vert^2}{\vert\chi(\dulR~\dulS,t)\vert^2} = 1.
\eea
This concludes the proof of Theorem I (a). It will become clear throughout this paper that, in many respects, the nuclear factor $\chi(\dulR~\dulS,t)$ can be 
viewed as a proper nuclear wavefunction. Like in the static case~\cite{GG05}, introducing the phase factor in Eq.~(\ref{eq:proof-Ia-phi}) allows 
$\chi(\dulR~\dulS,t)$ to have the correct antisymmetry if the nuclear subsystem contains identical fermionic nuclei.

 Next comes the question; what equations do
 $\Phi_{\dulR~\dulS}(\dulr~\duls,t)$ and $ \chi(\dulR~\dulS,t)$ satisfy? The answer entails the second part of Theorem I: 
\begin{widetext}
{\bf Theorem I (b)} {\it The wavefunctions $\Phi_{\dulR~\dulS}(\dulr~\duls,t)$ and $\chi(\dulR~\dulS,t)$ satisfy:} 
\ben
  \label{eq:exact_el_td}       
  \Bigl(\hat{H}_{el}(\dulr~\duls,\dulR~\dulS,t)-\epsilon(\dulR~\dulS,t)\Bigr)\Phi_{\dulR~\dulS}(\dulr~\duls,t)\\=i\partial_t \Phi_{\dulR~\dulS}(\dulr~\duls,t),
\een
\ben
 \label{eq:exact_n_td}
  \begin{split}
  \Bigl(\sum_{\alpha=1}^{N_n}\frac{1}{2M_\alpha}(-i\nabla_\alpha+{\bf A}_\alpha(\dulR~\dulS,t))^2 +\hat{V}_{ext}^n(\dulR,t) + \epsilon(\dulR~\dulS,t)\Bigr)\chi(\dulR~\dulS,t)=i\partial_t \chi(\dulR~\dulS,t),
\end{split}             
\een
{\it where the electronic Hamiltonian is}
\ben
\label{eq:e_ham_td}
\hat{H}_{el}(\dulr~\duls,\dulR~\dulS,t) = \hat{H}_{BO}(\dulr,\dulR,t)+\hat{V}\ext^e(\dulr,t) + \hat{U}_{en}^{coup}\left[\Phi_{\dulR~\dulS},\chi\right].
\een
{\it Here the electron-nuclear coupling potential $\hat{U}_{en}^{coup}\left[\Phi_{\dulR~\dulS},\chi\right]$, scalar potential $\epsilon(\dulR~\dulS,t)$, and vector potential ${\bf A}_\alpha(\dulR~\dulS,t)$ terms are}
\bea
\label{eq:exact_en_corr}
&&\hat{U}_{en}^{coup}\left[\Phi_{\dulR~\dulS},\chi\right]=\sum_{\alpha=1}^{N_n}\frac{1}{M_\alpha}\Big[\frac{(-i\nabla_\alpha-{\bf A}_\alpha(\dulR~\dulS,t))^2}{2} + \Big(\frac{-i\nabla_\alpha \chi(\dulR~\dulS,t)}{\chi(\dulR~\dulS,t)}+{\bf A}_\alpha(\dulR~\dulS,t)\Big)\cdot\left(-i\nabla_\alpha-{\bf A}_\alpha(\dulR~\dulS,t)\right)\Big]\\
\label{eq:exact_pes_td}
&&\epsilon(\dulR~\dulS,t) = \sum_{\duls}\left\langle\Phi_{\dulR~\dulS}(t) \right\vert\hat{H}_{el}((\dulr~\duls,\dulR~\dulS,t) - i \partial_t\left\vert \Phi_{\dulR~\dulS}(t)\right\rangle_{\dulr} \\
\label{eq:exact_BP_td}
&&{\bf A}_\alpha(\dulR~\dulS,t)=\sum_{\duls}\left\langle\Phi_{\dulR~\dulS}(t)\right\vert\left.-i\nabla_\alpha\Phi_{\dulR~\dulS}(t)\right\rangle_\dulr
\eea
{\it where $\langle ..|..|..\rangle_\dulr$ denotes an inner product over all spatial electronic variables only.} 
\end{widetext}

{\underline{\it Proof}} In order to derive the equations of motion for $\Phi_{\dulR~\dulS}(\dulr~\duls,t)$ and $\chi(\dulR~\dulS,t)$ we follow the strategy 
employed in the static case (see ref.~\cite{GG05}), i.e. we plug the product ansatz in the variational principle and search for the stationary point. Afterwards 
 we prove: if $\Phi_{\dulR~\dulS}(\dulr~\duls,t)$ and $\chi(\dulR~\dulS,t)$ are the solutions of Eqs.~(\ref{eq:exact_el_td}) and~(\ref{eq:exact_n_td}), then $\Phi_{\dulR~\dulS}(\dulr~\duls,t)\chi(\dulR~\dulS,t)$ is the solution of TDSE~(\ref{eq:tdse}). 
We begin the derivation by briefly reviewing Frenkel's stationary action principle as this is the key instrument to derive the equations of motion for each subsystem. 

The quantum mechanical action is defined as 
\ben
\label{eq:action} 
{\cal S}[\Psi,\Psi^*] = \int_{t_i}^{t_f} dt\langle\Psi\vert \hat{H} - i\partial_t\vert \Psi\rangle, 
\een
a functional of the time-dependent wavefunction $\Psi(t)$ and its 
complex conjugate. The equation of motion of the quantum system, the TDSE of Eq.~(\ref{eq:tdse}), is obtained by requiring the
variation of the action ${\cal S}$ with respect to all wavefunctions $\Psi(t)$ that satisfy the boundary condition
\ben
\label{eq:action-bound}
\delta \Psi(t_i) = \delta \Psi(t_f) = 0\;,
\een
to be stationary, i.e.,
\ben
\label{eq:action-stat} 
\delta_{\Psi^*} {\cal S} = 0. 
\een
Now we apply this general variational principle to our problem in the following way. We insert the product wavefunction in the action functional~(\ref{eq:action}), with Hamiltonian given by Eq.~(\ref{eq:H}), rewriting it as

\begin{widetext}
\bea
\label{eq:action-product} 
{\cal S}[\Phi_{\dulR~\dulS},\Phi^*_{\dulR~\dulS},\chi,\chi^*] 
   = \sum_{\duls,\dulS} \int_{t_i}^{t_f} dt \int d\dulR \int d\dulr \left[|\chi|^2 \Phi_{\dulR~\dulS}^{\ast} \left(\hat{H}_{BO} + \hat{V}^e_{ext}
 + \sum_{\alpha} \frac{-\na_{\alpha}^2}{2M_{\alpha}} - i\partial_t \right) \Phi_{\dulR~\dulS}\right.\nn\\
 + \left.|\Phi_{\dulR~\dulS}|^2 \chi^* \left(\sum_{\alpha}\frac{-\na_{\alpha}^2}
   {2M_{\alpha}}+ \hat{V}^n_{ext}- i\partial_t\right) \chi
 + |\chi|^2 \Phi_{\dulR~\dulS}^{\ast}\sum_{\alpha} \frac{1}{M_{\alpha}}(-i\nabla_\alpha\chi/\chi)\cdot(-i\na_{\alpha} \Phi_{\dulR~\dulS})\right]\;,\nn
\\
\eea
The equations of motion for $\Phi_{\dulR~\dulS}(\dulr~\duls, t)$ and $\chi(\dulR~\dulS, t)$ are obtained by requiring the action 
functional~(\ref{eq:action-product}) to be stationary with respect to variations of each wavefunction subject to the PNC~(\ref{eq:pnc}), i.e.,
\ben
\label{eq:action-stat-product}
\frac{\delta{\cal S}[\Phi_{\dulR~\dulS},\Phi^*_{\dulR~\dulS},\chi,\chi^*]}{\delta\Phi^*_{\dulR~\dulS}(\dulr~\duls,t)}= 0 \;\;\;{\rm and} \;\;\;
\frac{\delta{\cal S}[\Phi_{\dulR~\dulS},\Phi^*_{\dulR~\dulS},\chi,\chi^*]}{\delta\chi^*(\dulR~\dulS,t)}      = 0
\een
Variation of Eq.~(\ref{eq:action-product}) with respect to $\Phi_{\dulR~\dulS}^*(\dulr~\duls)$ leads to
\bea
\label{eq:variationphi1}
\vert\chi|^2 \left(\hat{H}_{BO} +\hat{V}_{ext}^e+ \sum_{\alpha} \frac{-\na_{\alpha}^2}{2M_{\alpha}}
-i\partial_t\right)\Phi_{\dulR~\dulS}+ 
\left[\chi^* \left(\sum_{\alpha} \frac{-\na_{\alpha}^2}{2M_{\alpha}}+\hat{V}_{ext}^n-i\partial_t\right) \chi\right] \Phi_{\dulR~\dulS} \nonumber\\
+ |\chi|^2 \left(\sum_{\alpha} \frac{1}{M_{\alpha}}(-i\nabla_\alpha\chi/\chi) \cdot(-i\na_{\alpha}\Phi_{\dulR~\dulS})\right) = 0\nonumber
\eea
Dividing the expression above by $|\chi|^2$ and rearranging yields:
\bea
\label{eq:variationphi2}
\left(\hat{H}_{BO}  +\hat{V}_{ext}^e+ \sum_{\alpha} \frac{-\na_{\alpha}^2}{2M_{\alpha}}-i\partial_t\right)\Phi_{\dulR~\dulS} + \sum_{\alpha} \frac{1}{M_{\alpha}}(-i\nabla_\alpha\chi/\chi)\cdot (-i\na_{\alpha} \Phi_{\dulR~\dulS})
= - \frac{(\sum_{\alpha} \frac{-\na_{\alpha}^2}{2M_{\alpha}} + \hat{V}_{ext}^n-i\partial_t) \chi} {\chi} \cdot \Phi_{\dulR~\dulS}\nonumber\\
\eea
Variation of Eq.~(\ref{eq:action-product}) with respect to $\chi^*$ yields
\bea
\label{eq:variationchi}
\left [\sum_{\duls}\int d \dulr \Phi_{\dulR~\dulS}^{\ast} \left (\hat{H}_{BO} +\hat{V}_{ext}^e
+ \sum_{\alpha} \frac{-\na_{\alpha}^2}{2M_{\alpha}}-i\partial_t\right)\Phi_{\dulR~\dulS}
\right]
 \chi &&+ \left[\sum_{\alpha} \frac{-\na_{\alpha}^2}{2M_{\alpha}}+\hat{V}_{ext}^n\right] \chi \nonumber\\
&&+ \left[\sum_{\alpha}\frac{1}{M_{\alpha}} (-i\nabla_\alpha\chi/\chi)
\cdot{\bf A}_{\alpha}\right] \chi = i\partial_t \chi\nn\\
\eea
\end{widetext}

where we enforced the PNC, and defined 
\bea
 {\bf A}_{\alpha} [\Phi_{\dulR~\dulS}]: = \sum_{\duls}\int d \dulr \Phi_{\dulR~\dulS}^{\ast}(\dulr~\duls)(-i\na_{\alpha}\Phi_{\dulR~\dulS}(\dulr~\duls)) \;.
\eea
This is a real-valued vector potential (see shortly). Inserting Eq.~(\ref{eq:variationchi}) on the RHS of Eq.~(\ref{eq:variationphi2}) 
leads, after some straightforward algebra, to
Eqs.~(\ref{eq:exact_el_td}-\ref{eq:exact_BP_td}).  The product
wavefunction Eq.~(\ref{eq:product-ansatz}), satisfying these
equations, therefore represents a stationary point of the action
functional~(\ref{eq:action-product}) under the PNC Eq.~(\ref{eq:pnc}).
To complete the proof, it remains to verify that if
$\Phi_{\dulR~\dulS} (\dulr~\dulS,t)$ satisfies
Eq.~(\ref{eq:exact_el_td}) and $\chi(\dulR~\dulS,t)$ satisfies
Eq.~(\ref{eq:exact_n_td}), then the product
$\Phi_{\dulR~\dulS}(\dulr~\duls,t)\chi(\dulR~\dulS,t)$ is an exact
solution of the TDSE. Approximate solutions of the TDSE may satisfy the
stationary action principle, if variations are taken over a limited
set of wavefunctions, e.g. the multi-configuration time-dependent
Hartree equations~\cite{MMC90} may be derived via the
Frenkel variational principle.  To dispel any possible doubts that the
product form of Eq.~(\ref{eq:product-ansatz}) subject to
Eq.~(\ref{eq:pnc}) is general, we now verify that our solution is exact
and not an approximation. Applying the product rule,
$
i\partial_t \Psi(\dulr~\duls,\dulR~\dulS,t) = \chi(\dulR~\dulS,t) i\partial_t \Phi_{\dulR~\dulS}(\dulr~\duls,t) + \Phi_{\dulR~\dulS}(\dulr~\duls,t)i\partial_t\chi(\dulR~\dulS,t)\;,          
$ and inserting ~Eqs.~(\ref{eq:exact_el_td}) and~(\ref{eq:exact_n_td}), we obtain
\begin{widetext}
\bea
\label{eq:implphi}
\nonumber
 \chi \left(i\partial_t \Phi_{\dulR~\dulS}\right) & = & \chi \left(\hat{H}_{BO}+\hat{V}_{ext}^e\right)\Phi_{\dulR~\dulS} + \chi \sum_{\alpha}^{N_n}\frac{(-i{\na}_{\alpha} 
 - {\bf A}_{\alpha})^2}{2M_{\alpha}}\Phi_{\dulR~\dulS} \\
 && + \chi \sum_{\alpha}^{N_n}\frac{(-i\nabla_{\alpha}\chi/\chi+{\bf A}_{\alpha})\cdot(-i{\na}_{\alpha} - {\bf A}_{\alpha})}{M_{\alpha}}\Phi_{\dulR~\dulS} - \chi \epsilon\Phi_{\dulR~\dulS} 
\eea

\bea
\label{eq:implchi}
 \Phi_{\dulR~\dulS}\left(i\partial_t \chi\right) & = & \Phi_{\dulR~\dulS} \sum_{\alpha}^{N_n}\frac{(-i\nabla_{\alpha} +{\bf A}_{\alpha}(\dulR~\dulS,t))^2}{2M_{\alpha}}\chi +\Phi_{\dulR~\dulS} \hat{V}^n_{ext} \chi + \Phi_{\dulR~\dulS} \epsilon \chi
\eea
\end{widetext}
Summing Eqs.~(\ref{eq:implphi}) and~(\ref{eq:implchi}) leads to the TDSE for the complete system and completes the proof that the wavefunctions satisfying Eqs.~(\ref{eq:exact_el_td}-\ref{eq:exact_BP_td})  do solve the TDSE exactly.

Alternatively, Eqs.~(\ref{eq:exact_el_td}-\ref{eq:exact_BP_td}) can be obtained by replacing $\Psi(\dulr~\duls,\dulR~\dulS,t)$, in the TDSE (\ref{eq:tdse}), by 
the product $\Phi_{\dulR~\dulS}(\dulr~\duls,t)\chi(\dulR~\dulS,t)$ and using the PNC~(\ref{eq:pnc}). The form of electron-nuclear coupling term, Eq.~(\ref{eq:exact_en_corr}), is the same as the static case (see ref.~\cite{GG05}). The exact TDPES, Eq.~(\ref{eq:exact_pes_td}), on the other hand is not simply the expectation
value of $\hat{H}_{el}$ but contains, in addition, the term $<\Phi_{\dulR~\dulS}|-i\partial_t\Phi_{\dulR~\dulS} >$. The appearance of this term is essential to 
ensure the form invariance of the Eqs.~(\ref{eq:exact_el_td}-\ref{eq:exact_BP_td}) under the gauge transformation~(\ref{eq:GT}) that will be discussed in Section IIB.

\subsection{Uniqueness of the electronic and nuclear wavefunctions}
\label{sec:thm2}
We now delve a little deeper into features of our exact factorization.
As briefly mentioned earlier, the factorization can be viewed in a
standard probabilistic setting~\cite{Hunter}: The square of the
molecular wavefunction can be viewed as a multivariate probability
distribution, that can be factorized into a marginal probability of a
set of variables (the nuclear coordinates) and a conditional
probability of the rest of the variables (the electronic coordinates,
conditionally dependent on the nuclear coordinates). In this sense we
identify $\chi(\dulR~\dulS,t)$ as the nuclear wavefunction (marginal
probability amplitude), and $\Phi_{\dulR~\dulS}(\dulr~\duls,t)$ as the
electronic wavefunction (conditional probability amplitude). An
equivalent formalism is to view, instead, the nuclear wavefunction as a
conditional probability amplitude depending parametrically on the
electronic coordinate, i.e. $\chi_{\dulr~\duls}(\dulR~\dulS,t)$, with the electronic wavefunction as the
marginal probability amplitude of the electronic coordinates, i.e. $\Phi(\dulr~\duls,t)$. We
choose to use the former decomposition however to later make natural
connections with the BO approach.
In this section we argue why we can view the probability amplitudes $\chi(\dulR~\dulS,t)$ and $\Phi_{\dulR~\dulS}(\dulr~\duls,t)$  as nuclear and electronic wavefunctions, and we will assign some meaning to the terms that arise in their equations of motion.

A first question that arises is: is this decomposition unique? We answer this in Theorem 2.

{\bf Theorem 2 (a)}  {\it Eqs.~(\ref{eq:exact_el_td}-\ref{eq:exact_BP_td})
are form-invariant up to within the gauge-like transformation:}
 \bea 
 \tilde{\Phi}_{\dulR~\dulS}(\dulr~\duls,t)&:=&e^{i\theta(\dulR~\dulS,t)}\Phi_{\dulR~\dulS}(\dulr~\duls,t)\nn\\
 \tilde{\chi}(\dulR~\dulS,t)&:=&e^{-i\theta(\dulR~\dulS,t)}\chi(\dulR~\dulS,t)
\label{eq:GT}
 \eea
\bea
\nonumber
       &&\mathbf{A}_\alpha(\dulR~\dulS,t)\rightarrow\tilde{\mathbf{A}}_\alpha(\dulR~\dulS,t)=\mathbf{A}_\alpha(\dulR~\dulS,t)+\nabla_{\alpha}\theta(\dulR~\dulS,t) \\
       &&\epsilon(\dulR~\dulS,t)\rightarrow\tilde{\epsilon}(\dulR~\dulS,t)=\epsilon(\dulR~\dulS,t) + \partial_t\theta(\dulR~\dulS,t) 
\label{eq:GTAeps}
\eea

    {\bf (b)} {\it The wavefunctions $\Phi_{\dulR~\dulS}(\dulr~\duls,t)$ and $\chi(\dulR~\dulS,t)$ are unique up to within the $(\dulR~\dulS,t)$-dependent phase transformation, Eq.~(\ref{eq:GT}).}

To prove part (a), simply substitute Eqs.~(\ref{eq:GT}) and~(\ref{eq:GTAeps}) into Eqs~(\ref{eq:exact_el_td})--(\ref{eq:exact_BP_td}).
Part (b) is readily shown by first assuming that $\Phi_{\dulR~\dulS}\chi$ and $\tilde{\Phi}_{\dulR~\dulS}\tilde{\chi}$ are two different representations of the 
 exact wave function $\Psi(\dulr~\duls,\dulR~\dulS,t)$ i.e.
 \ben
 \Psi(\dulr~\duls,\dulR~\dulS,t)=\Phi_{\dulR~\dulS}(\dulr~\duls,t)\chi(\dulR~\dulS,t) = \tilde{\Phi}_{\dulR~\dulS}(\dulr~\duls,t)\tilde{\chi}(\dulR~\dulS,t)
 \een
 Then
 \ben 
 \frac{\chi}{\tilde{\chi}}=\frac{\tilde{\Phi}_{\dulR~\dulS}}{\Phi_{\dulR~\dulS}} =: g(\dulR~\dulS,t)
 \een
 and
 \ben
 \vert \tilde{\Phi}_{\dulR~\dulS}(\dulr~\duls,t) \vert^2 = \vert g(\dulR~\dulS,t) \vert^2 \vert \Phi_{\dulR~\dulS}(\dulr~\duls,t) \vert^2.
 \een
 From Theorem 1, both $\tilde{\Phi}_{\dulR~\dulS}(\dulr~\duls,t)$ and $\Phi_{\dulR~\dulS}(\dulr~\duls,t)$ satisfy the PNC. Hence,
 \ben
 \sum_{\duls}\int d\dulr \vert \tilde{\Phi}_{\dulR~\dulS}(\dulr~\duls,t) \vert^2 = \vert g(\dulR~\dulS,t) \vert^2 \sum_{\duls}\int d\dulr \vert \Phi_{{\dulR~\dulS}}({\dulr~\duls},t) \vert^2
 \een
  and $\vert g(\dulR~\dulS,t) \vert^2 = 1$. Therefore, $g(\dulR~\dulS,t)$ must be equal to a purely $(\dulR~\dulS,t)$-dependence phase: 
 \ben
 g(\dulR~\dulS,t)=e^{i\theta(\dulR~\dulS,t)}.
 \een
This completes the proof of theorem 2. 

The interpretation of $\Phi_\dulR$ and $\chi$ as electronic and nuclear wavefunctions follows from the following observations. 
The probability density of finding the nuclear configuration $\dulR$ at time $t$, $\sum_{\duls}\int \vert\Psi(\dulr~\duls,\dulR~\dulS,t)\vert^2 d\dulr =\vert\chi(\dulR~\dulS,t)\vert^2$,  as can readily be shown by substituting the product wavefunction Eq.~(\ref{eq:product-ansatz}) into the left-hand-side and
using the PNC Eq.~(\ref{eq:pnc}).  Not only does $\chi(\dulR~\dulS,t)$ therefore yield the nuclear ($N_n$-body) probability density, we  shall see later in Section~\ref{sec:vecpot}, that it  also reproduces the exact nuclear
 ($N_n$-body) current-density.  
  The modulus-square of the electronic wavefunction,
$\vert\Phi_{\dulR~\dulS}(\dulr~\duls,t)\vert^2 = \vert\Psi(\dulr~\duls,\dulR~\dulS,t)\vert^2 / \vert \chi(\dulR~\dulS,t)\vert^2$ gives the conditional probability of finding the electrons at $\dulr$ with spin configuration $\duls$, given that the nuclear configuration is $\dulR~\dulS$.

Note that, strictly speaking, the definition of the conditional probability amplitude  $\vert\Phi_{\dulR~\dulS}(\dulr~\duls,t)\vert^2$ via
Eq.~(\ref{eq:proof-Ia-chi}),  only holds for non-zero marginal probabilities  $\vert\chi(\dulR~\dulS,t)\vert^2$. In the case the nuclear
density, and the full molecular wavefunction, have a node at some $\dulR_0$, the electronic wavefunction would be defined by taking a limit. 
However, it is actually very unlikely that the nuclear density has a node~\cite{Hunter2,CW78}. This can be seen by expanding the full
electron-nuclear wavefunction, $\Psi(\dulr~\duls,\dulR~\dulS,t)$, in terms of the BO-electronic states, as in Eq.~(\ref{eq:BO-Exp}). Then, the nuclear density can be
expressed as an infinite sum of non-negative terms:
\ben
 \label{eq:chi-nodeless}
 \vert \chi(\dulR~\dulS,t)\vert^2 = \sum_{j=1}^{\infty} \vert \chi^{BO}_j(\dulR~\dulS,t)\vert^2\;.
 \een
In general, it is extremely unlikely that every term in the summation
becomes zero at the same nuclear configuration $\dulR_0~\dulS_0$, unless dictated by symmetry~\cite{GG05} (see end of this section for a discussion on symmetry). Symmetry
dictated nodes likely lead to a finite, well-defined, value of $\vert\Phi_{\dulR~\dulS}(\dulr~\duls,t)\vert^2$ due to the linear behavior of the wavefunctions in the vicinity
of these nodes.

Eqs.~(\ref{eq:exact_el_td})-(\ref{eq:exact_BP_td}) determine the {\it
  exact} time-dependent molecular wavefunction, given an initial
state. As written, the nuclear equation is particularly appealing as a
Schr\"odinger equation with both scalar and vector-potential coupling
terms contributing effective forces on the nuclei including any
geometric phase effects. We call $\epsilon(\dulR~\dulS,t)$ and ${\bf
  A}(\dulR~\dulS,t)$ the {\it exact TDPES} and {\it exact time-dependent Berry
connection}, respectively. These two quantities, along with the
electron-nuclear coupling potential
$\hat{U}_{en}^{coup}[\Phi_{\dulR~\dulS},\chi]$, mediate the coupling
between the nuclear and the electronic degrees of freedom in a
formally exact way. The three sections in Section~\ref{sec:threeterms}
are each devoted to a closer study of these terms.

We conclude this section by discussing the symmetry properties of
$\chi(\dulR~\dulS,t)$ and $\Phi_{\dulR~\dulS}(\dulr~\duls,t)$: The
nuclear wavefunction $\chi(\dulR~\dulS,t)$ must preserve the symmetry
of the full electron-nuclear wavefunction
$\Psi(\dulr~\duls,\dulR~\dulS,t)$ with respect to exchange of
identical nuclei. This constrains the allowed gauge
transformation~(\ref{eq:GT})-(\ref{eq:GTAeps}). The electronic
wavefunction $\Phi_{\dulR~\dulS}(\dulr~\duls,t) = \Psi(\dulr~\duls,\dulR~\dulS,t)/\chi(\dulR~\dulS,t)$ is invariant under
any nuclear permutation because any fermionic sign cancels out between
the full molecular wavefunction and the nuclear wavefunction.
 
In the rest of the paper, we drop the spin indices $\dulS$ and $\duls$
for notational simplicity.

\subsection{Simple Illustration: the H atom in an electric field}
The example of the hydrogen atom in an electric field provides a
simple demonstration of our formalism. The Hamiltonian is
\ben
H = -\frac{1}{2M}\nabla_R^2 - \frac{1}{2}\nabla_r^2 - \frac{1}{\vert \bR - \br\vert} + (\br - \bR)\cdot {\bf E}(t)
\een
where $\br$ and $\bR$ are the  electron and proton coordinate respectively, ${\bf E(t)}$ is the applied electric field, and $M$ is the proton mass. 
The exact solution is known: in terms of the center of mass and relative coordinates, ${\bf R}_{\rm CM} = (\br + M\bR)/(M+1), {\bf u} = \br - \bR$, the problem is separable, and we have
\begin{equation}
\Psi({\bf R}_{\rm CM},{\bf u},t) =e^{i\bigl({\bf K} \cdot {\bf R}_{CM}-\frac{K^2}{2(M+1)}t\bigr)}\phi({\bf u},t)
\label{eq:HatomPsi}
\end{equation} 
where $\phi({\bf u},t)$ satisfies the following equation:
   \begin{equation}
     \left(-\frac{\nabla_u^2}{2\mu}-\frac{1}{u}+\mathbf{u}\cdot\mathbf{E}(t)\right)\phi(\mathbf{u},t)=i\partial_t\phi(\mathbf{u},t)
\label{eq:Hatom_phi}
   \end{equation} 
and
$\mu = M/(M+1)$ is the reduced mass. The full wavefunction, Eq.~(\ref{eq:HatomPsi}), represents  free-particle plane-wave motion in the center of mass coordinate, with
${\bf K}$  representing the total momentum of the system. The form of Eq.~(\ref{eq:HatomPsi}) suggests
one possible factorization for Eqs.~(\ref{eq:product-ansatz}) --(\ref{eq:pnc}) as:
  \bea
\nonumber
 \chi(\bR,t)&=&e^{i\left(\frac{-K^2t}{2(M+1)}+\frac{M}{(M+1)}\bf{K}\cdot\bR\right)}\\
     \Phi_{\bR}(\mathbf{r},t)&=&e^{i\frac{\mathbf{K}\cdot\mathbf{r}}{(M+1)}}\phi(\mathbf{r}-\bR,t)
\label{eq:Hatomdecomp}
\eea
with the exact Berry potential and TDPES given by 
\begin{equation}
     \mathbf{A}(\mathbf{R},t)=-i\int \phi^*(\mathbf{r}-\mathbf{R},t)\nabla_\mathbf{R}\phi(\mathbf{r}-\mathbf{R},t)d\mathbf{r} = 0
\label{eq:HatomA}
   \end{equation}
   \begin{equation}
     \epsilon(\mathbf{R},t)=\frac{K^2}{2(M+1)}+\mathbf{R}\cdot\mathbf{E}(t).
\label{eq:Hatomeps}
   \end{equation}
The vector potential, Eq.~(\ref{eq:HatomA}), is zero in the gauge implicit in our choice for Eqs.~(\ref{eq:Hatomdecomp}). This is easily confirmed by inserting 
Eqs.~(\ref{eq:Hatomdecomp}) in the nuclear equation~(\ref{eq:exact_n_td}), which reads for our problem,
\begin{equation}
\left(\frac{1}{M}\left(-i\nabla + {\bf A}\right)^2 - {\bf R}\cdot{\bf E}(t) +\epsilon({\bf R},t)\right)\chi({\bf R},t) = i\partial_t\chi({\bf R},t)
\label{eq:Hatom_neq}
\end{equation}
Eqs.~(\ref{eq:Hatomeps}) and~(\ref{eq:Hatom_neq}) show that, in this case
the role of the TDPES is to cancel out the external laser field in the nuclear equation, which is exactly as it should be. Only by this cancellation the nuclear 
motion can be a plane wave.

\section{The exact electron-nuclear coupling terms}
\label{sec:threeterms}
We now take a closer look at each of the three terms ${\bf A}(\dulR~\dulS,t)$,
$\epsilon(\dulR~\dulS,t)$,  and
$\hat{U}_{en}^{coup}[\Phi_{\dulR~\dulS},\chi]$, that mediate the
coupling between electron and nuclear dynamics exactly. In these three terms,  all of the non-adiabatic coupling effects of the Born-Oppenheimer expansion are effectively contained. 

\subsection{The time-dependent Berry connection}
\label{sec:vecpot} 
Eqs.~(\ref{eq:exact_el_td})-(\ref{eq:exact_BP_td}) demonstrate that a
Berry connection indeed appears in the exact treatment of coupled electron-ion dynamics, a question which was raised in the introduction. In this
section, we point out some properties of this object to help us
understand what it represents.

First, we show that the vector potential ${\bf A}_{\alpha}$ is real. 
Taking the gradient with respect to nuclear coordinates of the PNC (Eq.~(\ref{eq:pnc})), yields
\bea
0 &=&\na_{\alpha} \int d \dulr \Phi_{\dulR}^{\ast}(\dulr) \Phi_{\dulR} (\dulr)\nonumber\\
&=&2\mathrm{Re}  \int d \dulr \Phi_{\dulR}^{\ast}(\dulr)\na_{\alpha}\Phi_{\dulR}(\dulr)
\eea
(using the product rule). 
Comparing with the definition Eq.~(\ref{eq:exact_BP_td}), we readily conclude ${\bf A}_{\alpha}$ is real.

Second,  we insert Eqs.~(\ref{eq:proof-Ia-phi}) and~(\ref{eq:proof-Ia-chi}) into Eqs.~(\ref{eq:exact_BP_td}) to reveal the following expression for the vector potential:
\ben
  \label{eq:exact_vect}
  {\bf A}_\alpha(\dulR,t)=\frac{Im \left\langle\Psi(t)\right\vert\left.\nabla_\alpha\Psi(t)\right\rangle_\dulr}{\vert \chi(\dulR,t)\vert^{2}} - \nabla_{\alpha} S(\dulR,t)
\een
This shows that the vector potential is the difference of
paramagnetic nuclear velocity fields derived from the full and nuclear
wavefunctions. In fact, since $Im \left\langle\Psi(t)\right\vert\left.\nabla_\alpha\Psi(t)\right\rangle_\dulr$
is the true nuclear (many-body) current density,
Eq.~(\ref{eq:exact_vect}) implies that the gauge-invariant current
density, $Im (\chi^*\nabla_\alpha\chi)+\vert \chi\vert^{2}{\bf
  A}_\alpha$, that follows from the nuclear Hamiltonian in Eq.~(\ref{eq:exact_n_td}) does indeed
reproduce the exact nuclear current density~\cite{Manz}. As
discussed in the previous section, the solution $\chi(\dulR,t)$ of
Eq.~(\ref{eq:exact_el_td}) yields a proper nuclear many-body wavefunction:
Its absolute-value squared gives the exact nuclear ($N_n$-body) density
while its phase yields the correct nuclear ($N_n$-body) current density.
(The nuclear kinetic energy evaluated from $\chi(\dulR,t)$ does not equal the nuclear kinetic energy evaluated from the full molecular wavefunction, and their difference is determined by $U_{en}^{coup}$, as will be discussed in Section~\ref{sec:Uencorr}). 

Another interesting aspect of expression~(\ref{eq:exact_vect}) is that
it can help to shed light on the question of whether the exact Berry
potential produces a real effect or whether it can actually be
gauged away by a suitable choice of $\theta(\dulR,t)$ in
Eqs.~(\ref{eq:GT})-(\ref{eq:GTAeps}).  Provided the phase $S(\dulR,t)$
is spatially smooth, the last term on the right-hand-side of
Eq.~(\ref{eq:exact_vect}) can be gauged away so any true Berry
connection (that cannot be gauged away) must come from the first
term. In the conventional analyses of conical intersections, the phase
may not be smooth: for example, in the Herzberg and Longuet-Higgens
model~\cite{HLH63,M92}, the two (single-valued) nuclear wavefunctions
associated with a two-state conical intersection between traditional
BO surfaces, each have a phase $S = \pm \phi/2$, undefined at the
origin. This has a singular gradient, yielding a delta-function at the origin in the
curl of the vector potential, thus contributing a non-zero Berry phase.
Whether a similar effect occurs for the exact time-dependent nuclear
wavefunction remains to be explored. 
 When the exact $\Psi(t)$ is real-valued (e.g. for a
non-current-carrying ground state) then the first term on the
right-hand-side of Eq.~(\ref{eq:exact_vect}) vanishes and hence gives a vanishing contribution to the
exact Berry connection. Whether, and under which conditions, the full Berry connection (\ref{eq:exact_vect}) can be gauged away 
remains an open question at this point.

Finally, it is also instructive to express the  vector potential in terms of the BO electronic basis states of Section~\ref{sec:BOexp}. We first expand the electronic wavefunction:
\ben
\Phi_{\dulR}(\dulr,t) = \sum_{j=1}^{\infty} C_j(\dulR,t)\phi^j_{\dulR}(\dulr)
\label{eq:phi_exp}
\een
where orthonormality of the $\phi^j_\dulR$ (Eq.~(\ref{eq:BO-EN})) means
\ben
C_j(\dulR,t) = \int d\dulr\phi^{j*}_{\dulR}(\dulr)\Phi_{\dulR}(\dulr,t) \;.
\een
The PNC condition becomes
\ben
\sum_{j=1}^{\infty} \vert C_j(\dulR,t)\vert^2 = 1
\een
Inserting Eq.~(\ref{eq:phi_exp}) into Eq.~(\ref{eq:exact_BP_td}), and noting the definition of the non-adiabatic derivative couplings of Eq.~\ref{eq:NA-BO}, we obtain
\begin{widetext}
\ben
\label{eq:vector-BO}
{\bf A}_{\alpha}(\dulR,t) = \sum_{j=1}^{\infty} \Big(-iC^*_j(\dulR,t)\nabla_{\alpha}C_j(\dulR,t) + \vert C_j(\dulR,t)\vert^2 {\cal F}^{BO}_{jj,\alpha}(\dulR) + \sum_{l\neq j}^{\infty}C^*_l(\dulR,t)C_j(\dulR,t){\cal F}^{BO}_{lj,\alpha}(\dulR)\Big) 
\een
\end{widetext}
The exact Berry potential is thereby expressed as a linear combination
of the diagonal and off-diagonal BO derivative couplings. Any
gauge-invariant part of the Berry connection, that would give rise to
a non-zero Berry phase, arises from the part of Eq.~\ref{eq:vector-BO}
that has a non-zero curl.  In the case of a real-valued electronic
wavefunction, each of the three terms of Eq.~\ref{eq:vector-BO}
vanishes independently giving rise to a zero vector potential.

\subsection{The Time-Dependent Potential Energy Surface}
\label{sec:tdpes}
The time-dependent potential energy surface $\epsilon(\dulR,t)$
of Eq.~(\ref{eq:exact_pes_td}) provides an exact time-dependent
generalization of the adiabatic BO potential energy surface. As such,
it should prove to be a powerful interpretive tool for general
time-dependent problems. This will be explored in section IV. We now 
begin by analyzing the expression Eq.~(\ref{eq:exact_pes_td}) in a little more detail.

First, consider the expectation value of the
electron-nuclear coupling term, $\langle\Phi_{\dulR}\vert
\hat{U}_{en}^{coup}\vert\Phi_{\dulR}\rangle$ of
Eq.~(\ref{eq:exact_en_corr}) that appears in the TDPES. Only the first term of Eq.~(\ref{eq:exact_en_corr}) contributes to the expectation value: the second term goes to zero, due to the very last parenthesis, $\langle\Phi_{\dulR}\vert
-i\nabla_\alpha - {\bf A}_\alpha(\dulR,t)\vert\Phi_{\dulR}\rangle$, which vanishes due to the definition of the vector potential. So we have
\begin{widetext}
\bea
\epsilon(\dulR,t)& =& \Big(\langle\Phi_{\dulR}\vert \hat{H}_{BO} + \hat{V}_{ext}^{e}(\dulr,t) \vert \Phi_{\dulR}\rangle_{\dulr} - i\langle\Phi_{\dulR}\vert \partial_t \Phi_{\dulR}\rangle_{\dulr}+ 
\sum_{\alpha} \frac{\langle\Phi_{\dulR}\vert\left(-i\nabla_\alpha - {\bf A}_\alpha(\dulR,t)\right)^2 \vert\Phi_{\dulR}\rangle_{\dulr}}{2M_{\alpha}}\Big) \nonumber\\
&=& \Big(\langle\Phi_{\dulR}\vert \hat{H}_{BO} + \hat{V}^e_{ext}(\dulr,t) \vert \Phi_{\dulR}\rangle_{\dulr} - i\langle\Phi_{\dulR}\vert \partial_t \Phi_{\dulR}\rangle_{\dulr}+ 
\sum_{\alpha} \frac{\langle\nabla_{\alpha}\Phi_{\dulR}\vert \nabla_{\alpha}\Phi_{\dulR}\rangle_{\dulr}}{2M_{\alpha}}\Big) - \sum_{\alpha}\frac{{\bf A}^2_{\alpha}(\dulR,t)}
{2M_{\alpha}}
\label{eq:TDPES-BO}
\eea
\end{widetext}
where the second line results from expanding the square in the first, and making use of the definition of the vector potential. 

As we did for the vector potential, we now provide an expression
for the TDPES as an expansion over BO states. Inserting
Eq.~(\ref{eq:phi_exp}) into Eq.~\ref{eq:TDPES-BO} and performing a little
straightforward algebra, we obtain
\begin{widetext}
\bea
\epsilon(\dulR,t) = \sum_j \vert C_j(\dulR,t) \vert^2 V^j_{BO}(\dulR) 
+ \sum_{jl}C_j^*(\dulR,t)C_l(\dulR,t)\langle\phi^j_{\dulR}\vert \hat{V}^e_{ext}(\dulr,t) \vert \phi^l_{\dulR}\rangle_{\dulr} - \sum_j i C^*_j(\dulR,t)\partial_t C_j(\dulR,t) \nonumber\\
+ \sum_\alpha\frac{1}{2M_{\alpha}}\left(\sum_j \vert \nabla_{\alpha}C_j \vert^2 + \sum_{jl} C^*_jC_l \left(i\nabla_{\alpha}\cdot{\cal F}^{BO}_{jl,\alpha}-{\cal G}^{BO}_{jl,\alpha}\right) - 2 \sum_{jl} Im\big(C_l \nabla_{\alpha} C_j^* {\cal F}^{BO}_{jl,\alpha}\big) -  {\bf A}^2_{\alpha}(\dulR,t)\right)
\label{eq:TDPES-exp}
\eea
\end{widetext}
(the expansion of the last term ${\bf A}^2_{\alpha}$ may be obtained from Eq.~\ref{eq:vector-BO}). Notice that all the BO surfaces, as well as non-adiabatic couplings, are contained in the the exact TDPES. 

\subsection{Electron-Nuclear Correlation}
\label{sec:Uencorr}
The TDPES and Berry connection discussed in the previous two sections
directly determine the evolution of the nuclear wavefunction
(Eq.~(\ref{eq:exact_n_td})), containing the effect of coupling to the
electrons in an exact way.  The electron-nuclear coupling term
$\hat{U}_{en}^{coup}$ enters the nuclear equation indirectly via its
role in determining $\Phi_{\dulR}$ through Eq.~(\ref{eq:exact_el_td}) and (\ref{eq:e_ham_td}).
Eq.~(\ref{eq:exact_en_corr})
  expresses $\hat{U}_{en}^{coup}$ as a functional of the electronic
  and nuclear wavefunctions, and now we shall derive another
  expression for it that shows that it measures the difference between
  the nuclear kinetic energy evaluated from the full wavefunction and
  that evaluated on the nuclear wavefunction. We isolate
  the term involving $\hat{U}_{en}^{coup}$ in
  Eq.~(\ref{eq:exact_el_td}), and insert $\Phi_{\dulR} =
  \Psi/\chi$. This leads to:
  
\ben
\label{eq:Uen-exact1}       
\frac{\hat{U}_{en}^{coup}\Phi_{\dulR}}{\Phi_{\dulR}} = \frac{i\partial_t \Psi}{\Psi}-\frac{i\partial_t \chi}{\chi}-\frac{\hat{H}_{BO}\Phi_{\dulR}}{\Phi_{\dulR}}-\hat{V}\ext^e+\epsilon(\dulR,t) 
\een
Next we insert in Eq.~(\ref{eq:Uen-exact1}) the TDSE (\ref{eq:tdse}) and Eq.(\ref{eq:exact_n_td}), satisfied by $\Psi$ and $\chi$ to obtain
\ben
  \label{eq:Uen-exact2}       
 \frac{\hat{U}_{en}^{coup}\left[\Phi_{\dulR},\chi\right]\Phi_{\dulR}(\dulr,t)}{\Phi_{\dulR}(\dulr,t)} = \frac{\hat{T}_n\Psi}{\Psi} - \frac{\hat{\tilde{T}}_n \chi}{\chi}
\een
where 
\ben
\hat{\tilde{T}}_n = \sum_{\alpha=1}^{N_n}\frac{1}{2M_\alpha}(-i\nabla_\alpha+{\bf A}_\alpha(\dulR,t))^2 
\een
 Multiplying Eq.~(\ref{eq:Uen-exact2}) by $|\Phi_{\dulR}|^2|\chi|^2$ and integrating over all coordinates leads to:
\ben
\langle\Psi\vert\hat{T}_n\vert\Psi\rangle_{\dulr,\dulR} - \langle\chi\vert\hat{\tilde{T}}_n\vert\chi\rangle_\dulR = \int d\dulR \vert\chi(\dulR,t)\vert^2 \langle\Phi_{\dulR}\vert\hat{U}_{en}^{coup}\vert\Phi_\dulR\rangle_{\dulr} .
\een 

This means the nuclear kinetic energy evaluated from the full molecular wavefunction, and that evaluated via the expectation value of the nuclear kinetic energy operator in Eq.~(\ref{eq:exact_n_td}) on the nuclear wavefunction are not equal: their difference is given by the nuclear-density-weighted integral of the electron-nuclear coupling potential.

\section{Model of H$_2^+$ in a laser field}
\label{sec:H2plus}
\begin{figure}[h]
\centering
\includegraphics[width=1\columnwidth]{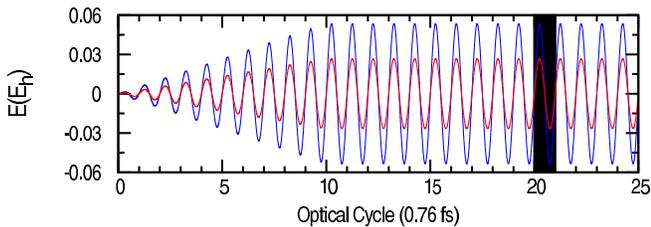}
\caption{$\lambda = 228$ nm laser field,  represented by $E(t) = E_0f(t)\sin(\omega t)$,
for two peak intensities, $I_1 = |E_0|^2 = 10^{14} W/$cm$^2$ and 
$I_2 = |E_0|^2 = 2.5 \times 10^{13} W/$cm$^2$. The envelope function $f(t)$ is chosen such that the field is linearly ramped from zero
to its maximum strength at $t=7.6$ fs and  thereafter held constant. The highlighted area represents the optical cycle that will be focussed on in later graphs.}
\label{fig:laser}
\end{figure}

In this section, we illustrate the usefulness of the TDPES using a
simple, numerically exactly solvable model: the $H_2^+$ molecular ion
subject to a linearly polarized laser field. By restricting the motion
of the nuclei and the electron to the direction of the polarization
axis of the laser field , the problem can be modeled with a 1D
Hamiltonian featuring ``soft-Coulomb'' interactions~\cite{JES88,LSWB96,VIC96,BN02,KLG04}: \bea
  \label{eq:H2+}
  \begin{split}
    \hat{H}(t) = & - \frac{1}{M}\frac{\partial^2}{\partial R^2} - \frac{1}{2\mu_e}\frac{\partial^2}{\partial x^2}+ \frac{1}{\sqrt{0.03+R^2}} + \hat{V}_l(x,t)\\
    &- \frac{1}{\sqrt{1+(x-R/2)^2}} - \frac{1}{\sqrt{1+(x+R/2)^2}} \\     
  \end{split}
\eea
where $R$ and $x$ are the internuclear distance and the electronic
coordinate as measured from the nuclear center-of-mass, respectively,
and the electronic reduced mass is given by $\mu_e=(2M)/(2M+1)$, $M$
being the proton mass. The laser field is represented by
$\hat{V}_l(x,t) = q_exE(t)$ where $E(t)$ denotes the electric field
amplitude and the reduced charge $q_e =(2M+2)/(2M+1)$.
One-dimensional soft-Coulomb atoms and molecules have proven extremely
useful in the study of strong-field dynamics since
they allow numerically accurate solutions to problems involving
correlated electron dynamics as well as correlated electron-nuclear
dynamics that would be computationally far more demanding for the full
three-dimensional atoms and molecules, while capturing the essential
physics of the latter, e.g. multi-photon ionization, above-threshold
ionization and dissociation, enhanced ionization,  non-sequential
double-ionization, high-harmonic generation, and non-BO effects (e.g. Refs.~\cite{VIC96,LL98,LKGE02,KLG04,KLEG01,BL05,BN02,CCZB96}).
We study
the dynamics of the model $H_2^+$ system under a $\lambda = 228$ nm ($5.4$eV) UV-laser pulse which is  represented by 
\ben
E(t) = E_0f(t)\sin(\omega t),
\een
with two peak intensities, $I_1 = |E_0|^2 = 10^{14} W/$cm$^2$ and $I_2 = |E_0|^2 = 2.5 \times 10^{13} W/$cm$^2$. With this frequency an energy that is about twice as much as the dissociation energy of the model molecule ($2.8782$eV) is achieved, so dissociation is expected. 
The envelope function $f(t)$ is chosen such that the field is linearly ramped from zero to its maximum strength at $t=T_{ramp}$ and thereafter held constant (Fig.~\ref{fig:laser}):
\ben
f(t) = \left \{ \begin{array} {lc}
  t/T_{ramp} & 0<t<T_{ramp}    \\
  1  &       T_{ramp}<t<T_{tot}
\end{array}\right. ,
\een
The rise-time was chosen as $T_{ramp} = 10 \tau$ while the total simulation time was $T_{tot} = 25 \tau$, where $\tau = \frac{2\pi}{\omega}$ denotes the optical cycle.
 
The same system and parameters were studied in Ref.~\cite{KLG04} where
the importance of electron-nuclear correlation was highlighted: a
two-configuration correlated ansatz for the time-dependent
electron-nuclear wavefunction was able to describe photodissociation
processes in many cases, while a simple uncorrelated Hartree product of an electronic and
a nuclear wavefunction almost always failed. In the present work we
analyse the dynamics via the numerically exact TDPES, finding it very useful in understanding and interpreting the motion. We note that the laser-field
does not couple directly to the nuclear relative coordinate $R$, but
only indirectly via the TDPES.

Starting from the exact ground-state as initial condition, we
propagate the TDSE numerically, using the second-order split-operator method~\cite{2nd-spo}, to obtain the full molecular
wavefunction $\Psi(x,R,t)$. As there is only one nuclear degree of
freedom (after separating off the center-of-mass motion), we can fix
the gauge in Eqs.~(\ref{eq:GT})-(\ref{eq:GTAeps}) such that the vector
potential~(\ref{eq:exact_vect}) vanishes identically. For one-dimensional problems this 
is always possible with the choice:
\ben
  \label{eq:exact_vect_H2+}
   \frac{d}{dR} S(R,t)=\frac{Im \int dx \Psi^*(x,R,t)\frac{d\Psi(x,R,t)}{dR}}{\vert \chi(R,t)\vert^{2}}. 
\een
 So we can calculate $S(R,t)$, the phase of the nuclear wavefunction, as well as $\vert \chi(R,t) \vert^2$, the nuclear density, from the computed exact time-dependent molecular wavefunction. Being equipped with the nuclear wave-function, $\chi(R,t)$ ($ = \vert \chi(R,t) \vert e^{iS(R,t)}$)
 , we then compute the TDPES by inverting the nuclear equation of motion~(\ref{eq:exact_n_td}).

 We will compare the exact dynamics with the following three approximations: (i) the usual
Ehrenfest approximation, where the nuclei are treated via classical
dynamics, evolving under the force $ - \nabla V_{Ehr} = -\nabla_\dulR
W_{nn}(\dulR) - \int d\dulr n(\br,t) \nabla_\dulR
W_{en}(\dulr,\dulR)$, with $n(\br,t)$ being the one-body electron
density , (ii) the ``exact-Ehrenfest'' approximation, which
substitutes the exact TDPES for the Ehrenfest potential $V_{Ehr}$ in
the usual Ehrenfest approach and, (iii) an uncorrelated approach, the
time-dependent Hartree (self-consistent field) approximation,
$\Psi_H(\dulr,\dulR,t)=\phi(\dulr,t)\chi(\dulR,t)$, where the
electronic part does not depend on $\dulR$ at all. This includes a
quantum treatment of the nuclei, but no electron-nuclear correlation.

\subsubsection{High intensity: $I_1 = 10^{14} W/$cm$^2$}
\begin{figure}[h]
\centering
\includegraphics[width=1\columnwidth]{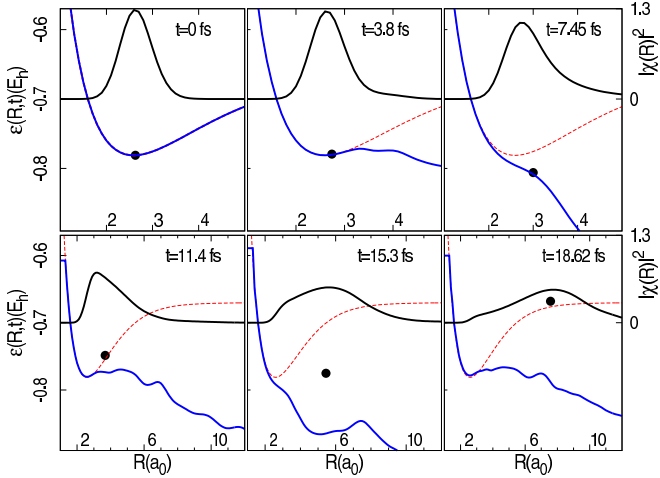}
\caption{Snapshots of the TDPES (blue solid lines) and nuclear density (black solid lines) at times indicated, for the H$_2^+$ molecule subject to the laser-field with the peak intensity $I_1 = 10^{14}$W/cm$^2$. The solid circles indicate the position and energy of the classical particle in the exact-Ehrenfest calculation. For reference, the ground-state BO surface (red dashed lines) is shown.}

\label{fig:tdpes1}
\end{figure}

The exact TDPES, along with the corresponding nuclear density,
 $\vert\chi(R,t)\vert^2$, are plotted in Fig.~\ref{fig:tdpes1} at six
 snapshots of time.  The initial TDPES lies practically on top of the
 ground-state BO surface, plotted in all the snapshots for
 comparison.

 The dissociation of the molecule is dramatically reflected in the
 exact TDPES, whose well flattens out, causing the nuclear density to
 spill to larger separations. Importantly, the tail of the TDPES
 alternately falls sharply and returns in correspondence with the
 field, letting the density out; the TDPES is the only potential
 acting on the nuclear system and transfers energy from the
 accelerated electron to the nuclei.

\begin{figure}[h]
\centering
\includegraphics[width=1\columnwidth]{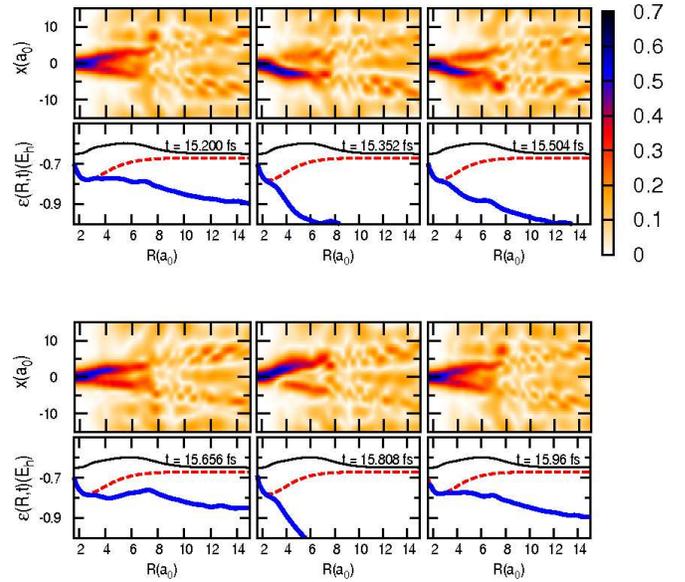}
\caption{Snapshots of the TDPES (blue lines), nuclear density (black) and the electronic conditional-density (color map) at times indicated during an optical cycle, for the H$_2^+$ molecule subject to the laser-field with the peak intensity $I_1 = 10^{14}$W/cm$^2$. For reference, the ground-state BO surface is shown as the red line.}
\label{fig:tdedens1}
\end{figure}

\begin{figure}[h]
\centering
\includegraphics[width=1\columnwidth]{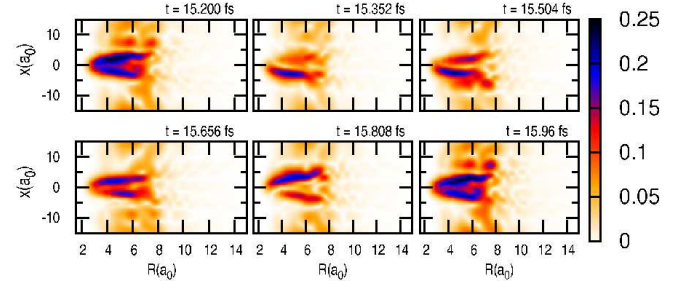}
\caption{Snapshots of the total electron-nuclear density at times indicated during an optical cycle, for the H$_2^+$ molecule subject to the laser-field with the peak intensity  $I_1 = 10^{14}$W/cm$^2$ .}
\label{fig:tdtotaldens1}
\end{figure}

In Figure~\ref{fig:tdedens1} we focus on six equally-spaced time
snap-shots during the optical cycle shaded in
Figure~\ref{fig:laser}. The lower panel shows the TDPES, with its
characteristic oscillations, along with the nuclear density as a
function of the internuclear coordinate, $\vert \chi(R,t)
\vert^2$. The upper panel shows a color map of the conditional
electronic probability density, $\vert \Phi_R(x,t)\vert^2$, i.e. the
probability of finding an electron at $x$ at a fixed nuclear
separation $R$. While at small internuclear distances (around and
below the equilibrium separation) the electron remains localized in
the middle between the two nuclei, at larger separations one clearly sees
the preferential localization of the electron density near the two nuclei, i.e. on 
one side or the other. At even larger separations we see streaks of ionizing electron
density in both directions. For the full story, we must multiply the
conditional probability density of the upper panels with the nuclear
density shown in the lower panel, to obtain the total electron-nuclear
density; this is shown in Figure~\ref{fig:tdtotaldens1},
indicating the probability of finding, at the time indicated, an
electron at position $x$ and the nuclear separation $R$.

\begin{figure}[h]
\centering
\includegraphics[width=1\columnwidth]{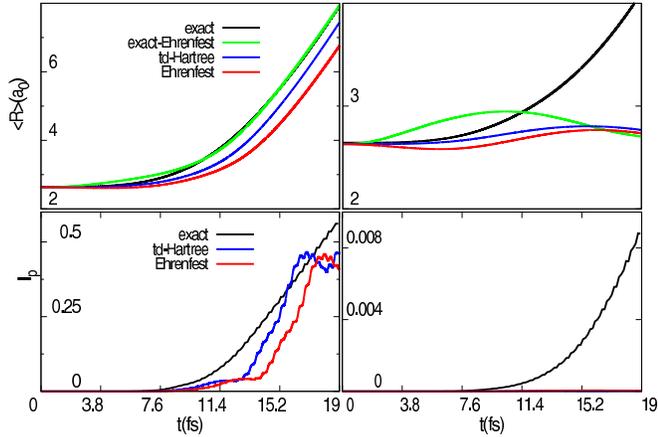}
\caption{Dissociation and ionization for intensity $I_1$ (left) and $I_2$ (right). Top panels: the internuclear separation $\langle R\rangle(t)$.  Lower panels: The ionization probability.}
\label{fig:R1}
\end{figure}

The top left-hand panel of Fig.~\ref{fig:R1} shows the expectation value of the internuclear
distance 
\ben
<\hat{R}> = \left\langle \Psi(t)\right\vert \hat{R} \left.\vert\Psi(t)\right\rangle,
\een
 along with the results from the three
approximate methods described earlier. The lower left-hand panel shows the
ionization probabilities. In principle, the latter requires projections of the full wavefunction on all continuum states which, in practice, are difficult to 
calculate. Alternatively, we use a geometrical concept~\cite{ABOX}, according to which the total ionization probabilities can be obtained from
\ben
\label{eq:ion-prob}
P_{ion}(t) = 1 -\int_{box_e} dx \left(\int dR |\Psi(t)|^2\right).
\een
The electrons leaving the ``electronic analyzing box'' ($box_e$) are thereby  identified with ionized electrons. The ionization box here was chosen to be
$|x|\leq10$. The internuclear distance together with the ionization probability support a Coulomb-explosion
interpretation of the dissociation: first, the system begins to
ionize, then the nuclei begin to  rapidly move apart under their mutual Coulomb
repulsion increasingly sensed due to weaker screening by the reduced
electron density. 
Turning now to the approximations, we observe that  all the
methods yield dissociation and some ionization.  The expectation value of the internuclear
distance in Fig.~\ref{fig:R1}, demonstrates that among all the
approximate calculations employed here, the exact-Ehrenfest is most
accurate. Referring back to  Figure~\ref{fig:tdpes1}: the solid circles indicate the classical nuclear position and energy of a particle driven by the exact-Ehrenfest force. One can see that it rapidly picks up kinetic energy above the TDPES, supporting the fact that the nuclear dissociation mechanism is an essentially classical one in this case.
The exact-Ehrenfest calculation even does better than TD-Hartree which
treats the protons quantum mechanically, thus showing the overarching importance
of electron-nuclear correlation in this case.

\begin{figure}[h]
\centering
\includegraphics[width=1\columnwidth]{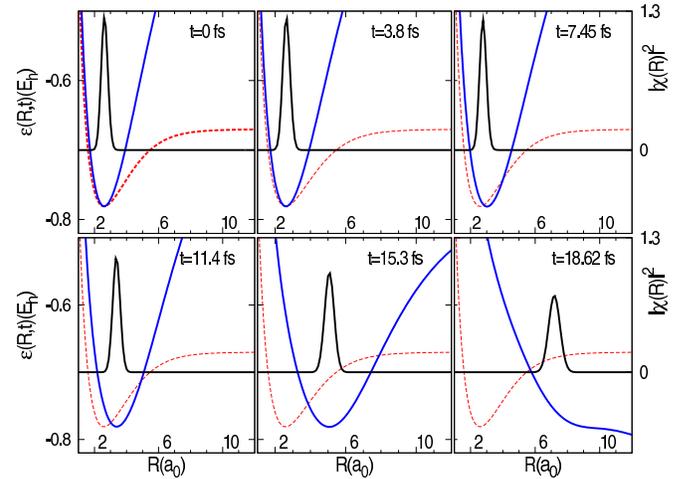}
\caption{Snapshots of the time-dependent Hartree nuclear-potential (blue lines) and nuclear density (black) at times indicated, for the H$_2^+$ molecule subject to the laser-field with the peak intensity  $I_1 = 10^{14}$W/cm$^2$. For reference, the ground-state BO surface is shown as the red line.}
\label{fig:tdh1}
\end{figure}

In fact, the Hartree description is worse than it may seem from just
looking at the internuclear separation in
Fig.~\ref{fig:R1}. In
Figure~\ref{fig:tdh1} we plot the time-dependent
Hartree potential energy surface and Hartree nuclear-density.  Both
are dramatically different from the exact TDPES and exact nuclear
density of Figure~\ref{fig:tdpes1}. At the initial time, the
Hartree potential is reasonably good near equilibrium but poor at
large separations~\cite{KLG04}: this is a consequence of the
conditional electron probability being independent of the nuclear
coordinate, and therefore only yielding a realistic result where the
energy is optimized, which is at equilibrium separation. As time
evolves the minimum of the Hartree surface moves out and begins to
widen, cradling the nuclear density, which more or less retains its
Gaussian shape, unlike the exact density; only at larger times does
the surface open out.

\subsubsection{Lower intensity: $I_2 = 2.5 \times 10^{13} W$/cm$^2$} 
\begin{figure}[h]
\centering
\includegraphics[width=1\columnwidth]{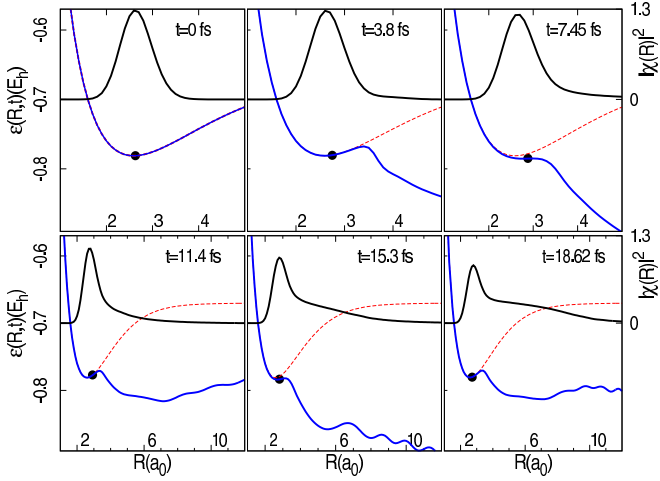}

\caption{Snapshots of the TDPES (blue) and nuclear density (black) at times indicated, for the H$_2^+$ molecule subject to the laser-field with the peak intensity $I_2 = 2.5 \times 10^{13}$W/cm$^2$. The solid circles indicate the position and energy of the classical particle in the exact-Ehrenfest calculation. For reference, the ground-state BO surface (dashed red) is shown.}
\label{fig:tdpes2}
\end{figure}

We now consider the dynamics under a field of weaker intensity.
Figure ~\ref{fig:tdpes2} plots the TDPES, whose tail displays similar
oscillations as in the higher intensity case. The nuclear density appears
to leak out to larger separations, although more slowly than in the previous case; indeed from the right panels in Fig.~\ref{fig:R1}, we
see that the exact calculation leads to dissociation.  However,
Fig.~\ref{fig:R1} (upper right panel)also shows that none of the approximations
dissociate, in contrast to the previous case. The Hartree and Ehrenfest methods also  show negligible ionization, compared to the exact case; but even in the exact case the ionization probability is very small, indicating a different mechanism of dissociation than in the stronger field case.
It may
be at first surprising that the exact-Ehrenfest calculation does not
dissociate the molecule, given that it is based on the exact TDPES,
however an examination of classical dynamics in the TDPES of
Fig.~\ref{fig:tdpes1} can explain what is happening. The solid dot in
Fig.~\ref{fig:tdpes1} indicates the classical position and energy, and
we see that it is always trapped inside a well in the TDPES, that remains at all
times.
This suggests that tunneling is the leading
mechanism for the dissociation:
a classical particle can only oscillate inside the well, while a quantum particle may tunnel out, as indeed
reflected in Fig.~\ref{fig:R1}.  Although the tail has similar
oscillations as for $I_1$, this does not lead to dissociation of
classical nuclei due to the barrier; the TDPES in this case transfers
the field energy to the nuclei via tunneling. Although the
exact-Ehrenfest calculation shows a larger amplitude of oscillation than the others,
it ultimately cannot tunnel through the barrier. 

\begin{figure}[h]
\centering
\includegraphics[width=1\columnwidth]{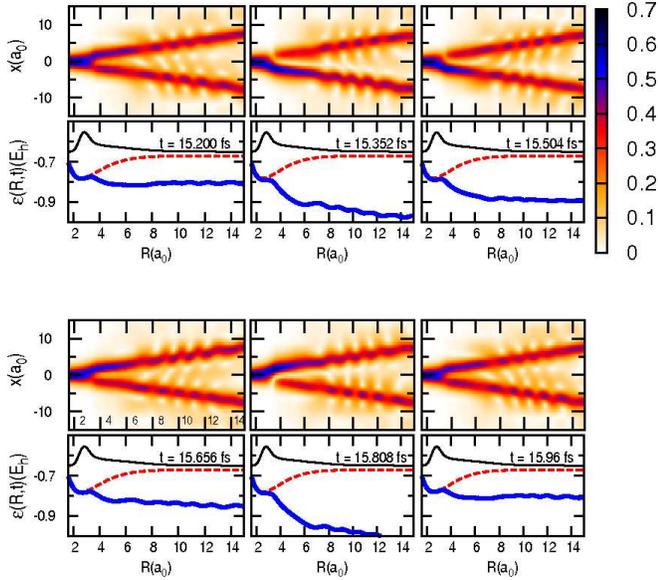}
\caption{Snapshots of the TDPES (blue lines), nuclear density (black) and the electronic conditional-density (color map) at times indicated during an optical cycle, for the H$_2^+$ molecule subject to the laser-field with the peak intensity  $I_2 = 2.5 \times 10^{13}$W/cm$^2$ . For reference, the ground-state BO surface is shown as the dashed red line.}

\label{fig:tdedens2}
\end{figure}

\begin{figure}[h]
\centering
\includegraphics[width=1\columnwidth]{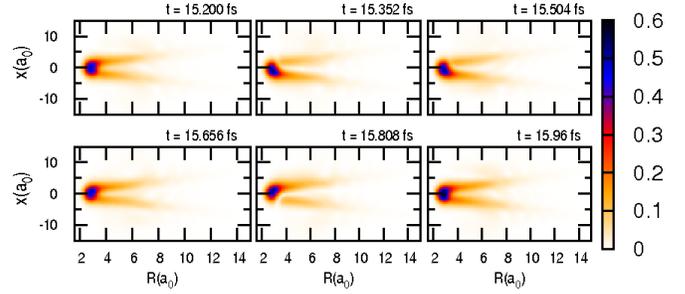}
\caption{Snapshots of the total electron-nuclear density at times indicated during an optical cycle, for the H$_2^+$ molecule subject to the laser-field with the peak $I_2 = 2.5 \times 10^{13}$W/cm$^2$.}
\label{fig:tdtotaldens2}
\end{figure}

As in the previous case, we plot in the top panels of
Fig.~\ref{fig:tdedens2} the electronic conditional density
$\vert\Phi_R(x,t)\vert^2$ over one optical cycle, while the lower
panels illustrate again the opening and closing of the TDPES as the
field oscillates. Like in the previous case, for small $R$ near
equilibrium, the electron density is localized in between the nuclei,
while for larger $R$, there is some polarization towards one side or
the other. To get the full picture, one must multiply the top panels
by the nuclear density $\vert\chi(R,t)\vert^2$, to obtain the total
electron-nuclear probability density, shown in
Figure~\ref{fig:tdtotaldens2}. It is evident in this graph that there
is much less ionization than in the previous case, and the dissociation is slower.

\begin{figure}[h]
\centering
\includegraphics[width=1\columnwidth]{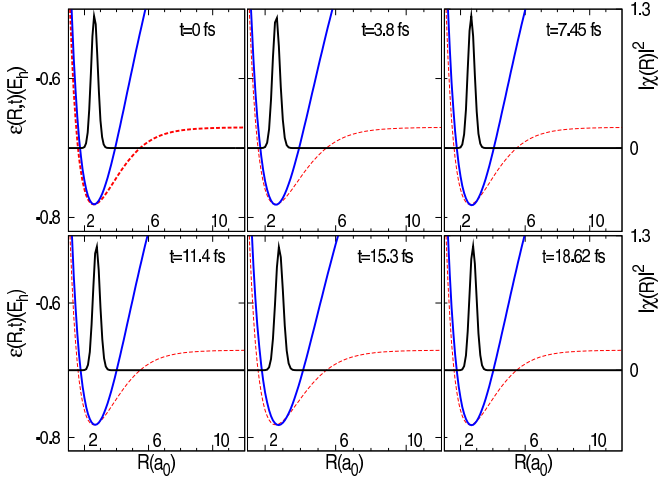}
\caption{Snapshots of the time-dependent Hartree nuclear-potential (blue lines) and nuclear density (black) at times indicated, for the H$_2^+$ molecule subject to the laser-field with the peak intensity  $I_1 = 10^{14}$W/cm$^2$  $I_2 = 2.5 \times 10^{13}$W/cm$^2$. For reference, the ground-state BO surface is shown as the dashed red line.}
\label{fig:tdh2}
\end{figure}

Although the Hartree approximation treats the nuclei quantum
mechanically, and therefore allowing tunneling in principle, tunneling and
dissociation do not actually occur. The reason for this is clear from
the shape of the Hartree potential, plotted in Fig.~\ref{fig:tdh2}:
the Hartree potential essentially retains its initial shape at all
times, making very small oscillations near the equilibrium
separation. As in the more intense field case, this is due to its
uncorrelated treatment of the electron-nuclear system: the electronic
wavefunction at any nuclear configuration is always the same, and is
best at equilibrium since initially it is determined by
energy-optimization, from where it does not deviate far, due to the
weak field strength. Unlike in the stronger field case, the Hartree
surface never opens out.  Dissociation via tunneling requires both a
quantum mechanical description of the nuclei and an adequate
accounting of electron-nuclear correlation. 

We do not expect the TDPES to be so different from the BO surfaces in
all cases. For example, in the case of field-free vibrational dynamics
of the H2+ molecule, where we start with a nuclear wavepacket
displaced from equilibrium on the ground BO surface, we find the TDPES
follows closely the BO surface throughout.  The non-adiabatic
couplings are weak in this case. The TDPES for field-free dynamics in
other systems with stronger non-adiabatic couplings will be published
elsewhere~\cite{Aetal}.

The purpose of comparing the exact results with these methods (TD-Hartree, Ehrenfest and exact-Ehrenfest) 
was primarily to support the conclusions drawn from the exact TDPES regarding 
the dissociation mechanisms. An interesting question is how well do the more accurate 
approximate PES's proposed recently (e.g. Ref~\cite{KSTK04}) compare with the exact TDPES; 
this will be investigated in the future.

\section{Conclusions}
In this paper, we have shown that there 
exists a rigorous factorization of the exact molecular wavefunction into
a nuclear wavefunction and electronic wavefunction, each of which
retains the usual probabilistic meaning.
The exact nuclear $N_n$-body density is
$\vert\chi(\dulR,t)\vert^2$ while $\vert\Phi_{\dulR}(\dulr,t)\vert^2$
represents the conditional probability of finding the electrons at
$\dulr$, given the nuclear configuration $\dulR$.
Equations~(\ref{eq:exact_el_td})---(\ref{eq:exact_BP_td}) are the
equations of motion that the electronic wavefunction and nuclear
wavefunction satisfy, and show explicitly how the electronic and
nuclear systems are exactly coupled.  These equations enable the
time-dependent potential energy surface (Eq. (\ref{eq:exact_pes_td}))
and the time-dependent Berry connection (Eq. (\ref{eq:exact_BP_td})) to be 
defined as rigorous concepts, and we have discussed some general properties of them, and
of the electron-nuclear coupling operator Eq.~(\ref{eq:exact_en_corr}).

 The example of the one-dimensional H$_2^+$ molecule in an oscillating
 electric field, solved numerically accurately, demonstrated that the
 TDPES is a powerful tool to analyze and interpret different types of
 dissociation processes. By studying the shape and evolution of the
 TDPES, comparing classical dynamics in this exact potential to the
 exact quantum dynamics, we were able to distinguish whether the dissociation
 proceeded via nuclear tunneling or more directly in
 Coulomb-explosion. For this example, the TDPES is the only potential
 determining the nuclear dynamics, exactly containing the coupling
 with electronic dynamics. The example demonstrated the importance of
 capturing both quantum effects in nuclear motion and electron-nuclear
 coupling; the Hartree approach, for example, despite treating the
 nuclei quantum mechanically, was unable to capture dissociation via
 tunneling as the shape of its potential surface was completely wrong. Thus, 
 for exactly solvable systems, the TDPES, and in more general cases than the 
 one studied here, the geometric phase, can be very useful interpretative tools for dynamics. 
The calculation of a TDPES has quite some history in the strong-field community, 
and several possible definitions of TDPES have been proposed in the literature. The crucial 
point of our work is that it provides a {\it unique} definition of TDPES 
(unique up to within a gauge transformation): If one wants the TD many-body
Schr\"odinger equation~(\ref{eq:exact_n_td}) to give the correct N-body density and current 
density of the nuclei, then the scalar potential and the vector potential 
{\it must} be given by eq.~(\ref{eq:exact_pes_td}) and~(\ref{eq:exact_BP_td}). There is no choice apart from the gauge. 
That means that with any advanced technique that yields the TD 
molecular wavefunction $\Psi(\dulr,\dulR,t)$ one can evaluate the TDPES and 
Berry potential by first calculating the factors from Eqs.~(\ref{eq:proof-Ia-phi})--(\ref{eq:proof-Ia-chi}) and then 
evaluating the TDPES and Berry potential from Eqs.~(\ref{eq:exact_pes_td})--(\ref{eq:exact_BP_td}). 

From a practical point of view, Eqs.~(\ref{eq:exact_el_td})-(\ref{eq:exact_BP_td}) are not easier to solve than the time-dependent Schr\"odinger equation for the full electron-nuclear system.
Rather they form the rigorous starting point for
making approximations, especially for the systematic development of
semiclassical approximations. In the large-nuclear mass limit, the electronic 
equation reduces to Cederbaum's time-dependent
BO approximation~\cite{C08,AMG10}.  Taking the classical limit for the nuclei in the
large-mass limit, one retrieves the Ehrenfest equations with Berry
potential~\cite{AMG10} (see also~\cite{K07,ZW06}). Treating the nuclei classically but retaining
their finite mass, one finds corrections to the Ehrenfest equations
that better account for non-adiabatic transitions~\cite{AAG12}. A
direction for future research is to capture some nuclear quantum
effects by a semiclassical or quasiclassical procedure~\cite{Ciccotti, Miller}, built on the
exact foundational equations presented here. Another direction would be to use the formalism as a 
possible starting point to develop electron-nuclear correlation functionals in a density-functionalized 
version of the electron-nuclear problem~\cite{KG01}. A promising route is to develop a time-dependent 
generalisation of the optimized effective potential scheme proposed in~\cite{GG05}. 

{\it Acknowledgments:}
Partial support from the National Science
Foundation (CHE-1152784) (NTM), from the Deutsche Forschungsgemeinschaft (SFB 762) and from the European Commission (FP7-NMP-CRONOS) is gratefully acknowledged.


\begin{thebibliography}{99}

\bibitem{AMG10} A. Abedi, N. T. Maitra, E. K. U. Gross, Phys. Rev. Lett. {\bf 105}, 123002 (2010). 

\bibitem{BK03}
Bandrauk, A.D., and H. Kono, ``Molecules in intense laser fields: nonlinear multiphoton
spectroscopy and near-femtosecond to sub-femtosecond (attosecond) dynamics'', in Advances in MultiPhoton Processes and Spectroscopy, vol. 15, edited by S.H. Lin, A.A.
Villaeys, and Y. Fujimura, pp. 147--214 (World Scientific, Singapore, 2003)

\bibitem{Marangos04}
Marangos, J.P., ``Molecules in a strong laser field'', in Atoms and Plasmas in Super-
Intense Laser Fields, edited by D. Batani, C. J. Joachain, and S. Martellucci, SIF Conference
Proceedings, vol. 88, pp. 213--243 (Societ`a Italiana di Fisica, Bologna, 2004)

\bibitem{Kling04}
M. F. Kling et al. Science {\bf 312}, 246 (2006). 

\bibitem{DP07}
W. R. Duncan and O. V. Prezhdo, 
Annu. Rev. Phys. Chem. {\bf 58}, 143, (2007). 

\bibitem{Rozzi11}
C. Rozzi et al., unpublished.

\bibitem{CCZB96}
S. Chelkowski et al., Phys. Rev. A {\bf 54}, 3235 (1996); 


\bibitem{Martin07}
F. Martín et al., Science {\bf 315}, 629 (2007).

\bibitem{paramonov05}
G.K. Paramonov, Chem. Phys. Lett. {\bf 411}, 350 (2005).

\bibitem{HBNS06}
A. P. Horsfield et al., Rep. Prog. Phys. {\bf 69}, 1195 (2006).

\bibitem{Martinez}
M. Ben-Nun et al., J. Phys. Chem. A {\bf 104}, 5161 (2000).

\bibitem{TIW96p}
A. D. McLachlan, Mol. Phys. {\bf 8}, 39 (1964); J. C. Tully, Faraday Discuss. {|bf 110}, 407 (1998); J. C. Tully, J. Chem. PHys. {\bf 93}, 1061 (1990);
M. Thachuk, M.Yu Ivanov, D.M. Wardlaw, J. Chem. Phys. {\bf 105}, 4094 (1996); 
M.A.L. Marques et al., Comp. Phys. Commun. {\bf 151}, 60 (2003). 

\bibitem{TTRF08}
E. Tapavicza et al. J. Chem. Phys. {\bf 129}, 124108 (2008). 
\bibitem{PDP09}
O. V. Prezhdo, W. R. Duncan, and V. V. Prezhdo, Prog. Surf. Sci. {\bf 84}, 30 (2009). 
\bibitem{BS81}
A. D. Bandrauk and M. Sink, J. Chem. Phys. {\bf 74}, 1110 (1981).

\bibitem{KSTK04}
H. Kono et al., Chem. Phys. {\bf 304}, 203 (2004). 

\bibitem{KSIV11}
F. Kelkensberg et al. Phys. Chem. Chem. Phys. {\bf 13}, 8647 (2011). 

\bibitem{C08}
L.S. Cederbaum, J. Chem. Phys. {\bf 128}, 124101 (2008).

\bibitem{Hunter}
G. Hunter, Int. J. Quant. Chem. {\bf 9}, 237 (1975).

\bibitem{GG05} 
Nikitas I. Gidopoulos, E. K. U. Gross, arXiv:cond-mat/0502433.

\bibitem{Berry}
M.V. Berry, Proc. R. Soc. A {\bf 392}, 45 (1984).

\bibitem{K03}
B. K. Kendrick, J. Phys. Chem. A {\bf 107}, 6739 (2003). 

\bibitem{R00}
R. Resta, J. Phys.: Condens. Matter {\bf 12}, R107 (2000). 

\bibitem{BALB10}
F. Bouakline, S. C. Althorpe, P. Larregaray, and L. Bonnet, Mol. Phys. {\bf 108}, 969 (2010). 

\bibitem{A06}
S. Althorpe, J. Chem. Phys. {\bf 124}, 084105 (2006). 

\bibitem{M92}
C. A. Mead, Rev. Mod. Phys. {\bf 64}, 51 (1992). 

\bibitem{BH54}
M. Born, K. Huang, Dynamical Theory of Crystal Lattices, Oxford University, New York, 1954.

\bibitem{CBBO}
L. S. Cederbaum, ``Born-Oppenheimer Approximation and Beyond'', in Advanced Series in Physical Chemistry, Vol. 15, {\it Conical intersections: 
electronic structure, dynamics and spectroscopy}, edited by Wolfgang Domcke, David Yarkony and Horst Köppel, pp. 3--40 (World Scientific, Singapore, 2004)

\bibitem{Baer}
M. Baer, {\it Beyond Born-Oppenheimer: Conical Intersections and Electronic Nonadiabatic Coupling Terms}, (John Wiley and Sons, 2006)

\bibitem{Hunter2}
G. Hunter, Int. J. Quant. Chem. {\bf XIX}, 755-761 (1981).

\bibitem{CW78}
J. Czub and L. Wolniewicz, Mol. Phys. 36, 1301 (1978).

\bibitem{Manz}
I. Barth et al., Chem. Phys. Lett. {\bf 481}, 118 (2009). 

\bibitem{JES88}
J. Javanainen, J. Eberly, and Q. Su, Phys. Rev. A {\bf 38}, 3430 (1988).

\bibitem{LSWB96}
D.G. Lappas, A. Sanpera, J.B. Watson, K. Burnett, P.L. Knight, R. Grobe, J.H. Eberly, J. Phys. B 29,  L619 (1996). 

\bibitem{VIC96}
D.M. Villeneuve, M.Y. Ivanov, and P.B. Corkum, Phys.
Rev. A. {\bf 54}, 736 (1996).

\bibitem{BN02}
 A. Bandrauk and H. Ngyuen, Phys. Rev. A. {\bf 66}, 031401(R) (2002). 

\bibitem{KLG04}
T. Kreibich, R. van Leeuwen, and E. K. U. Gross, Chem. Phys. {\bf 304}, 183 (2004). 

\bibitem{LL98}
D. G. Lappas and R. van Leeuwen, J. Phys. B: At. Mol. Opt. Phys. {\bf 31}, L249 (1998). 

\bibitem{LKGE02}
M. Lein et al., Phys. Rev. A {\bf 65}, 033403 (2002).

\bibitem{KLEG01}
T. Kreibich et al. Phys, Rev. Lett. {\bf 87}, 103901 (2001).

\bibitem{BL05}
A. D. Bandrauk and H. Lu, Phys. Rev. A. {\bf 72}, 023408 (2005). 


\bibitem{2nd-spo}
J. A. Fleck, J. R. Morris, and M. D. Feit, Appl. Phys. A 10, 129 (1976)

\bibitem{Aetal}
A. Abedi et al. , unpublished (2012).




\bibitem{MMC90}
H. D. Meyer, U. Manthe, and L. S. Cederbaum, Chem. Phys. Lett. {\bf 165}, 73 (1990). 

\bibitem{HLH63}
G. Herzberg and H. C. Longuet-Higgens, Discuss. Faraday Soc. {\bf 35}, 77 
(1963).

\bibitem{KG01}
T. Kreibich and E.K.U. Gross, Phys. Rev. Lett. 86, 2984 (2001).

\bibitem{ABOX}
K.C. Kulander , Phys. Rev. A. {\bf 35}, 445 (1987). 

\bibitem{K07}
V. Krishna, J. Chem. Phys. {\bf 126}, 134107 (2007).

\bibitem{ZW06}
Qi Zhang, Biao Wu, Phys. Rev. Lett. {\bf 97}, 190401 (2006).

\bibitem{AAG12}
A. Abedi, F. Agostini, and E. K. U. Gross, submitted (2012).


\bibitem{Ciccotti}
R. Kapral, G. Ciccotti, J. Chem. Phys. {\bf 110}, 8919 (1999).

\bibitem{Miller}
W. H. Miller, J. Phys. Chem. A {\bf 113}, 1405 (2009).


\end{thebibliography}
\end{document}